\documentclass[%
 reprint,
nofootinbib,
 amsmath,amssymb,
 aps,
]{revtex4-2}

\usepackage{graphicx}
\usepackage{dcolumn}
\usepackage{bm}
\usepackage{enumerate}
\usepackage{xspace}
\newcommand{\tc}{\textdegree\xspace}
\usepackage{xcolor} 

\begin{document}


\title{Simulating droplet adhesion on superhydrophobic surfaces}
\author{Pawan Kumar}
\email{kumar.pawan.cot@gmail.com}
\affiliation{%
 Department of Chemical Engineering, University of Melbourne, Parkville, Melbourne, 3010, Victoria, Australia}%
\author{Joseph D. Berry}%
 \email{berryj@unimelb.edu.au}
\affiliation{%
 Department of Chemical Engineering, University of Melbourne, Parkville, Melbourne, 3010, Victoria, Australia}%


\date{\today}

\begin{abstract}
A quasi-static numerical model is proposed to simulate the adhesion, compression, and subsequent detachment of a microliter droplet from a superhydrophobic surface composed of chemically homogeneous pillars arranged in a periodic fashion, replicating a typical force probe microscopy experiment. We observe that as the droplet is pulled away from the surface, the net vertical force varies in a typical sawtooth manner with peculiar peaks and troughs, characteristic of the surface. The force first reaches a maximum before the droplet detaches from the surface with a comparatively lower force. The force variation predicted by the numerical model is in good agreement with experimental results. 
We also studied the effect of evaporation on the variation in the adhesion force by simulating an evaporating droplet on a superhydrophobic surface. For an evaporating droplet, the numerically predicted maximum and detachment force magnitudes are in good agreement with those obtained experimentally when we take into account the change in the droplet weight as it evaporates. 
The proposed method will be useful for the quantitative analysis and design of a variety of superhydrophobic surfaces and will pave the way for more accurate surface characterization based on droplet adhesion force measurements. 

\end{abstract}

\maketitle


\section{Introduction}
\label{sec:intro}
Since the discovery of the Lotus leaf effect \cite{barthlott1997purity}, a variety of surfaces have been developed to cater to specific wetting applications, for example, superhydrophobic water repellent surfaces \cite{bhushan2011natural,nishimoto2013bioinspired}, sticky superhydrophobic surfaces (SSHS) \cite{feng2008petal,jiang2024sticky}, and Slippery Liquid-Infused Porous Surfaces (SLIPS) \cite{manna2015fabrication,peppou2020life}. These surfaces are used in many industrial and engineering systems, for example  in anti-icing \cite{farhadi2011anti,kreder2016design}, anti-fouling \cite{amini2017preventing,sunny2016transparent}, and drag reducing \cite{bocquet2011smooth,xu2020superhydrophobic} applications. Sticky superhydrophobic surfaces are used in energy harvesting \cite{zhang2019liquid,li2019spontaneous} and self-cleaning surfaces \cite{bird2013reducing}, while SLIPS are used in applications requiring high droplet mobility \cite{villegas2019liquid}. The key to the functioning of these surfaces is the ability to control the mobility of a liquid droplet in contact with the surface. When a droplet is placed on a superhydrophobic surface, the surrounding air is trapped within the surface crevices and partially supports the droplet. The droplet is in contact with the surface asperities only at certain places, and the smaller the number of these contact points, that is, the roughness area fraction ($\phi$), the higher the droplet mobility. The mobility of the droplets can be quantified by measuring the advancing contact angle ($\theta_{\rm{a}}$) and the contact angle hysteresis (CAH). High $\theta_{\rm{a}}$ ($>150$\textdegree) and low CAH ($<10$\textdegree) are often associated with high mobility \cite{quere2005non,roach2008progess,daniel_probing_2023} (with the exception of SLIPS and Slippery Omniphobic Covalently Attached Liquid surfaces (SOCAL) \cite{wang2016covalently} that exhibit low $\theta_{\rm{a}}$ and low CAH \cite{daniel_probing_2023}). 

Although CAH and receding contact angle characterize a superhydrophobic surface, the physical parameters of interest, such as the adhesion force ($F$) and the lateral friction force ($F_{\rm{fric}}$) between the surface and the droplet, cannot be directly determined. 
For example, Furmidge's equation \cite{furmidge1962studies} relates $F_{\rm{fric}}$ to the droplet parameters as
\begin{equation}
    F_{\rm{fric}}=k\sigma_{\rm{la}}w\left(\cos\theta_{\rm{r}} - \cos\theta_{\rm{a}} \right),
    \label{eqn:fric_force}
\end{equation}
where $w$ is the width of the droplet, $\theta_{\rm{r}}$ is the receding contact angle and $\sigma_{\rm{la}}$ is the interfacial tension of the liquid-air interface. $k$ is a geometrical prefactor that depends on the shape of the droplet that must be assumed depending on the specific case \cite{stern2025furmidge}. \citeauthor{mchale2022friction} \cite{mchale2022friction} used a solid-liquid friction law similar to solid-solid friction (Amonton's laws) to predict the friction force of droplets on different surfaces using the experimental data of \citeauthor{gao2018drops} \cite{gao2018drops}. 

The adhesion force between a droplet and a surface can also be indirectly inferred using the following equation \cite{koh2024quantifying},
\begin{equation}
    F = \left(\pi R^2 \right) \Delta p - 2 \pi R \sigma_{\rm{la}} \sin \theta_{\rm{r}},
    \label{eqn:adh_force}
\end{equation}
where $R$ is the radius of the droplet base and $\Delta p$ is the Laplace pressure difference across the liquid-air interface. Clearly, to relate the contact angle values to the adhesion or friction forces, additional parameters such as the droplet base radius, the Laplace pressure difference, and a proportionality constant ($k$) are also required. 
For this reason, direct force measurement is preferred when parameters such as adhesion or lateral friction forces are to be determined. In addition, Contact Angle Goniometry (CAG) \cite{allred2017wettability,huhtamaki2018surface}, which is the most widely used technique to measure contact angles, is less sensitive to changes in surface wettability compared to direct force measurements \cite{daniel_mapping_2019}. Other scenarios where direct force measurements may be preferred over CAG are discussed 
in other relevant studies \cite{kumar2025adhesion, daniel_mapping_2019,daniel_quantifying_2020,liimatainen2017mapping}. 

Direct force measurement 
can be broadly categorized into two approaches: (a) force probe microscopy \cite{hokkanen_forcebased_2021, liimatainen2017mapping, pilat_dynamic_2012, ko2020single} and (b) atomic force microscopy (AFM) \cite{daniel_mapping_2019, daniel_quantifying_2020, shi_measuring_2015}. Both methods operate on similar principles but differ in the size of the droplet used. Force probe microscopy employs millimeter-sized droplets \cite{daniel2017oleoplaning,daniel2019hydration,backholm2020water,hinduja2022scanning}, whereas AFM uses droplets with diameters on the order of tens of micrometers \cite{daniel_mapping_2019,escobar2017measuring,shi2016long,daniel_quantifying_2020}. 
To measure the lateral friction force, in force probe microscopy, a millimeter-sized droplet is attached to the end of an elastic material (such as a glass capillary) with a known spring constant. The capillary deflects as the droplet moves relative to the surface. This deflection, measured via a laser or camera, is 
used to calculate the lateral friction force \cite{hinduja2022scanning}. The velocity of the relative motion is adjustable, enabling friction force measurements as a function of velocity. 
To measure the adhesion force, the droplet is attached to a disk \cite{liimatainen2017mapping}, ring \cite{samuel2011study} or the tip of a needle \cite{livi2025characterization}, and brought into contact with the surface using a motorized stage. A sensitive microbalance measures the force during the interaction between the droplet and the  surface.

Both the lateral friction forces \cite{hinduja2022scanning,backholm2020water,lepikko2025droplet,backholm2024toward} and the adhesion forces \cite{liimatainen2017mapping,samuel2011study,nagy2024determination,hokkanen_forcebased_2021,dong2018superoleophobic} depend on the surface topography and chemical nature. However, during lateral force measurement, the droplet is asymmetric with simultaneous advancing and receding interfaces. As a consequence the dynamics of the contact line (CL) are complex in comparison to  adhesion force measurements, where the entire CL is either receding or advancing. In addition, the droplet wetting and spreading behavior on superhydrophobic surfaces is predominantly governed by the receding contact line dynamics as the advancing contact angle stays constant, approaching 180\tc \cite{oner2000ultrahydrophobic, priest2007asymmetric,choi2009modified,dorrer2007contact,gauthier2013role,dorrer2006advancing, iliev2016contact,kusumaatmaja2007modeling,rivetti2015surface,mognetti2010modeling, priest2009asymmetric, reyssat2009contact, yeh2008contact, kwon2010cassie}  (\citeauthor{schellenberger2016water} using inverted laser scanning confocal microscopy demonstrated this in \cite{schellenberger2016water}). Therefore, direct adhesion force measurement is the most suitable for analysis of superhydrophobic surfaces. 

The majority of the work on studying the adhesion forces on superhydrophobic surfaces focuses on developing experimental techniques \cite{hokkanen_forcebased_2021, liimatainen2017mapping, pilat_dynamic_2012, gao2018drops, daniel_mapping_2019, daniel_quantifying_2020, shi_measuring_2015, daniel2017oleoplaning,daniel2019hydration,backholm2020water,hinduja2022scanning, daniel_mapping_2019,shi_measuring_2015,shi2016long,daniel_quantifying_2020,lepikko2025droplet,backholm2024toward,samuel2011study,nagy2024determination,dong2018superoleophobic} with relatively fewer analytical or numerical treatments.  \citeauthor{butt2008capillary} \cite{butt2008capillary} proposed a formalism to calculate the capillary forces between rough solids connected by a capillary bridge. However, surface roughness was treated statistically and the effect of pillar geometry was not considered. This model is extremely useful for understanding the physics of capillary bridge formation and the effect of surface roughness, but it cannot be directly used to predict forces between a droplet and a pillared superhydrophobic surface. 
\citeauthor{baret2006wettability} \cite{baret2006wettability} studied the stability of a droplet attached to a pipette in contact with a flat surface by solving the Young-Laplace equation \cite{daniel2023droplet}, but the study was limited to flat surfaces only and cannot be used as such to analyze the adhesion on pillared surfaces. \citeauthor{sadullah2024predicting} \cite{sadullah2024predicting} used Lattice Boltzmann (LB) simulations to model a Centrifugal Adhesion Balance (CAB) system \cite{tadmor2008centrifugal} and predicted droplet detachment for a variety of surfaces. However, only the effect of the receding contact angle ($\theta_{\rm{r}}$) on the detachment force was studied without considering the variation of the force with time and the associated dynamics of the CL. \citeauthor{chen2017direction} \cite{chen2017direction}
used LB simulations to calculate the adhesion forces between a pillared surface and a droplet studying the effect of the wetting transition between the Wenzel and Cassie-Baxter states, although the  variation of force (with time or the position of the platform on which the droplet is placed) and the effect of pillar area fraction on the droplet adhesion were not considered. \citeauthor{sudersan2023method} \cite{sudersan2023method} used energy minimization simulations to calculate the variation in the adhesion force between a droplet attached to the tip of a cantilever and a flat surface as the tip is moved away from the surface. To the best of our knowledge, there is no numerical model available in the literature that can simulate the variation in the adhesion force between a droplet and a pillared superhydrophobic surface, replicating a typical force probe or atomic force microscopy experiment. 

In this work, we present a numerical method to simulate the force probe microscopy using a microliter droplet on a superhydrophobic surface composed of cylindrical pillars arranged in a square pattern. We used this model to calculate the variation in the adhesion force as a function of time (in simulation units; discussed in detail in \S \ref{sec:force}) and compare it with  experimental results. 
The numerical model can simulate both force probe and atomic force microscopy, hereafter referred to as droplet probe microscopy. In addition, we also discuss the effect of evaporation on the variation in the adhesion force on superhydrophobic surfaces by simulating an evaporating droplet. Although cylindrical pillars arranged in a square pattern are considered in this work, the model is valid for arbitrary pillar geometries and distributions. The numerical model presented is useful for predicting the adhesion force on a variety of superhydrophobic surfaces and can also be used to develop a drop probe microscopy tool for surface characterization.

\section{Physical system and the Numerical Model}
\label{sec:method}

In this section, we discuss the approach for simulating the variation in net vertical force between a microliter droplet and a superhydrophobic surface composed of micrometer-sized pillars, capturing a typical droplet probe microscopy experiment. The volume of droplets used in this study is small; therefore, the Bond number for the system $Bo=\rho g R_d^2/\sigma_{\rm{la}}$, where $\rho$ is the liquid density, $g$ is the acceleration due to gravity and $R_d$ is the radius of the droplet, is much smaller than unity. For example, for a 1.6 $\mu$L water droplet in air, $Bo=0.07\ll1$. Hence, the effect of gravity has been neglected in the present study.  Although we discuss the method in the context of a millimeter sized droplet, the approach can be used to simulate smaller droplets as long as the surface roughness length scale $l_{\rm{c}}\ll R$. 
The dynamics of the system are simulated as a series of equilibrium droplet morphologies 
(see \cite{kumar2024numerical} for details on the method) obtained by minimizing the total free energy of the droplet. 
The physical model is presented first followed by the numerical method used to capture the physics.
\subsection{Physical Model}
\label{sec:physical_model}

\begin{figure*}
    \centering
    \includegraphics[width=\linewidth]{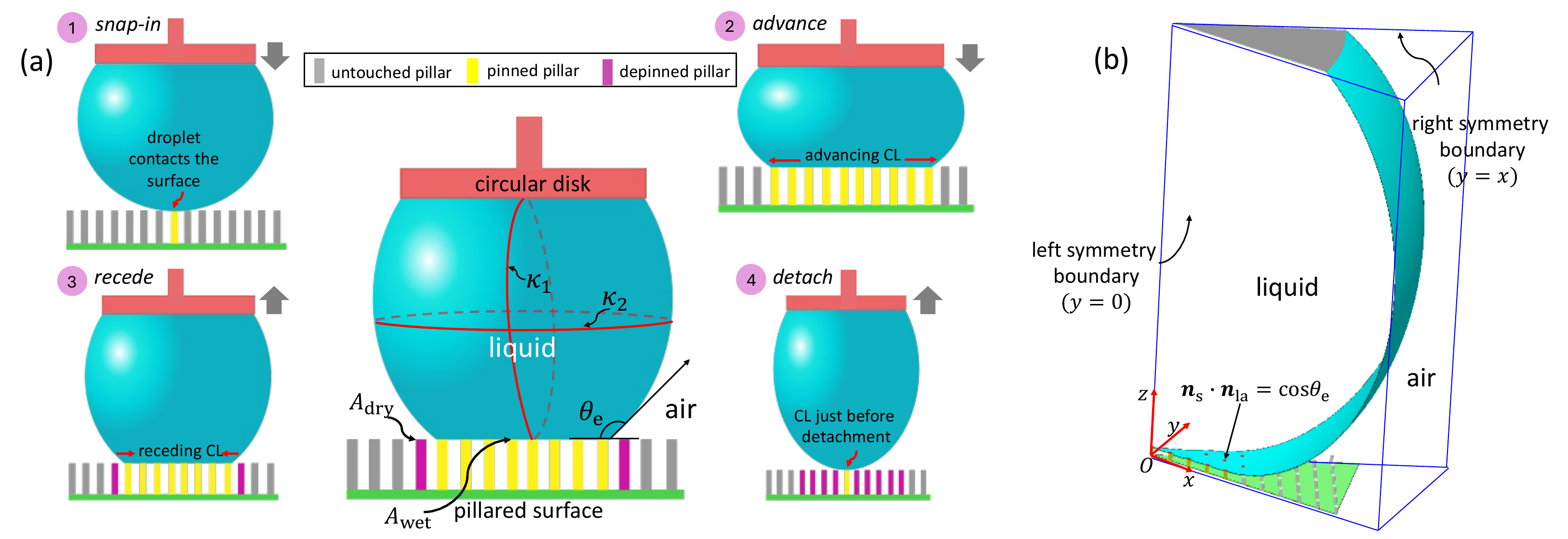} 
    \caption{(a) Pictorial representation of a typical droplet probe microscopy simulation. The method is depicted in 4 key steps (\textit{approach} stage is not shown here): 
    (1) \textit{Snap-in} - the droplet contacts the surface, (2) \textit{Advance} - the droplet is pushed against the surface by lowering the disk, (3) \textit{Recede} - the droplet is pulled away from the surface by moving the disk in the opposite direction, and (4) \textit{Detach} - the droplet detaches from the surface. (b) Schematic of the simulation domain. Due to symmetry, only one-eighth of the full droplet is simulated to reduce the computational time. 
    The pillars in contact with the droplet are shown in yellow, the pillars from which the CL has depinned are shown in magenta and the pillars untouched by the liquid are shown in gray/white colors.
    Symmetry boundary conditions is applied on the left and right walls of the domain. Young's angle boundary condition is applied at the contact line. In equilibrium, the droplet maintains the condition of Young's angle boundary condition at the CL and the liquid-air interface being a surface of constant mean curvature ($\kappa_1+\kappa_2$ being constant).}
    \label{fig:cv}
\end{figure*}
In a typical droplet probe microscopy experiment, a small droplet is attached at the end of a disk, a needle, or a cantilever (Fig. \ref{fig:cv}). The droplet is fixed relative to the disk but can move with the disk relative to the surface, which is kept on a movable or stationary stage. The entire process comprises five key stages, namely \textit{approach}, \textit{snap-in}, \textit{advance}, \textit{recede}, and \textit{detach}, as depicted in Fig. \ref{fig:cv}a.
During the \textit{approach} stage, the droplet is brought closer to the surface either by moving it downward or by moving the surface upward via a movable stage. In the simulation, we model droplet/surface interaction by the downward and upward movement of the disk during the \textit{advance} and \textit{recede} stages, respectively. 
When the droplet first contacts the surface, it wets the top of the contact pillar, forming a CL that is pinned at the top edge of the pillar. The spread of the droplet on the top of the pillar takes place at a relatively fast capillary speed ($v_{\rm{CV}}$), which is roughly 1.6 m/s for the water-air system (and surface roughness of the order of approximately 10 $\mu$m) \cite{kumar2024experimental}. This rapid movement of the interface is the \textit{snap-in}. As the disk/stage continues to move, the droplet gets compressed and the interface moves radially outward, wetting the tops of more pillars. The advancement of the CL over the surface is the \textit{advance} stage. At the start of the \textit{recede} stage, the movement of the disk/stage is reversed. This results in a receding of the interface that culminates in the detachment of the droplet, which is the \textit{detach} stage.


The total interfacial energy ($E^{'}$) of the system changes as the droplet moves through these stages. During the \textit{approach} stage,
\begin{equation}
 E^{'} = \sigma_{\rm{la}}A_{\rm{la}} + \sigma_{\rm{sl}}A_{\rm{disk}} + \sigma_{\rm{sa}}A_{\rm{surf}}, 
 \label{eqn:energy1}
\end{equation}
where $\sigma_{\rm{la}},\sigma_{\rm{sl}},\sigma_{\rm{sa}}$ are the interfacial tension of the liquid-air, solid-liquid \footnote{The interfacial tension of the solid-liquid interface between the droplet and the material of the disk will be different from that between the droplet and the pillared surface, but for simplicity both are considered as the same.}, and solid-air interfaces, respectively, $A_{\rm{la}}$ is the area of the liquid-air interface, $A_{\rm{disk}}$ is the area of the disk which is in contact with the droplet, and $A_{\rm{surf}}$ is the total surface area of the pillared surface. When the droplet contacts the surface (\textit{snap-in}), a portion of the liquid-air interface is converted into the solid-liquid interface, while the same amount of solid-air interfacial area is removed. The total interfacial energy of the system when the droplet is in contact with the surface can be written as 
\begin{equation}
 E^{'} = \sigma_{\rm{la}}A_{\rm{la}} + \sigma_{\rm{sl}}A_{\rm{disk}} + \sigma_{\rm{sl}}A_{\rm{wet}} + \sigma_{\rm{sa}}A_{\rm{dry}},
 \label{eqn:energy2}
\end{equation}
where $A_{\rm{wet}}$ and $A_{\rm{dry}}$ are the wet and dry areas of the pillared surface ($A_{\rm{dry}}=A_{\rm{surf}}-A_{\rm{wet}}$). Rearranging 
Eq. (\ref{eqn:energy2}),
\begin{equation}
 E = \sigma_{\rm{la}}\left(A_{\rm{la}} - \cos\theta_{\rm{e}}A_{\rm{wet}}\right).
 \label{eqn:enegy3}
\end{equation}
Here $E = E^{'} - \sigma_{\rm{sl}}A_{\rm{disk}} -\sigma_{\rm{sa}}A_{\rm{surf}}$ and $\theta_{\rm{e}}$ is the equilibrium contact angle, which is equal to the Young angle for an ideal (perfectly flat and chemically homogeneous) surface. Since $\sigma_{\rm{sl}}$, $\sigma_{\rm{sa}}$, $A_{\rm{disk}}$, and $A_{\rm{surf}}$ are constants for a particular system, the change in the interfacial energy based on $E^{'}$ or $E$ is the same. Henceforth, we use $E$ to calculate any changes in the interfacial energy of the system. During each stage, the droplet takes a shape for which it has a minimum energy (Eq. (\ref{eqn:enegy3})). Note that in Eq. (\ref{eqn:enegy3}), $\theta_{\rm{e}}$ is replaced with the inherent advancing ($\theta_{\rm{A}}$) or receding ($\theta_{\rm{R}}$) contact angle to account for the presence of inherent contact angle hysteresis in real surfaces \cite{raj2014high, he2004contact} (see also \S \ref{sec:interface_dynamics}). During the \textit{advance} and \textit{recede} stages, the CL advances and recedes on the surface respectively, and therefore $A_{\rm{wet}}$ increases and decreases respectively. $A_{\rm{wet}}$ is zero when the droplet has not yet contacted the surface and when the droplet has detached from the surface (assuming that no micro-droplets are generated, which is a reasonable assumption, as the surfaces considered in the present work are not superhydrophilic, which may result in interfacial necking and generation of small droplets as observed by \citeauthor{darvish2026water} \cite{darvish2026water}). For the intermediate states, $A_{\rm{wet}}$ changes depending on the position of the CL on the surface.

At any stage, the droplet will be in equilibrium in contact with the solid surface if it satisfies two conditions: 
\begin{enumerate}[I. ]
    \item the droplet has minimum total energy (Eq. (\ref{eqn:enegy3}))
    \item the liquid-air interface locally intersects the solid surface at Young's angle, that is, $\boldsymbol{n}_{\rm{s}} \cdot \boldsymbol{n}_{\rm{la}}= \cos\theta_{\rm{e}}$
\end{enumerate}.
Here, $\boldsymbol{n}_{\rm{s}}$ and $\boldsymbol{n}_{\rm{la}}$ are the unit normal vectors for the solid surface (directed outward) and the liquid-air interface (directed towards the air), respectively. During the \textit{advance} and \textit{recede} stages, there may be instances when these two conditions are not satisfied simultaneously, resulting in the interface not being in equilibrium. Before we discuss such scenarios, it should be noted that in equilibrium, the droplet surface can also be represented by a surface of constant mean curvature, which follows from the solution of the Euler-Lagrange equation under a constant volume constraint (see \citeauthor{de2003capillarity} \cite{de2003capillarity} pp. 29-30). This means that the total curvature ($\kappa$) at every point on the interface is the same, that is, $\kappa=\kappa_1+\kappa_2=c$. Here, $\kappa_1$ and $\kappa_2$ are the principal curvatures in the vertical (meridional) and horizontal (azimuthal) planes respectively (see Fig. \ref{fig:cv}a), and $c$ is a constant depending on the position of the disk relative to the surface. 

We now briefly discuss the kinematics of the CL by considering the liquid-air interface as a surface of constant mean curvature. Note that the droplet shape obtained by minimizing the total interfacial energy (Eq. (\ref{eqn:enegy3})) is the same as that obtained by solving for a surface of constant mean curvature under the given kinematic constraints. As the disk is lowered down during the \textit{advance} stage, the system kinematics dictate that the meridional curvature increases. If condition II, Young's angle boundary condition, is satisfied at the contact line, the droplet finds a new equilibrium morphology with new values of meridional and azimuthal curvatures. However, as the disk continues to move downwards, the liquid-air surface eventually intersects neighboring pillars. This results in the CL advancing, and a new equilibrium droplet morphology where condition II is now being satisfied on the advanced CL. Therefore, the CL advances in a rolling motion, bending towards and then touching a new set of pillars in the advancement direction, which is typical of Cassie-Baxter wetting state \cite{schellenberger2016water}.

The \textit{recede} stage of the droplet is more complex. During the \textit{recede} stage, the disk is moved in the opposite direction, gradually pulling the droplet away from the surface. The system kinematics dictates that the meridional curvature of the droplet decreases, along with the azimuthal curvature and the Laplace pressure inside the droplet. As the droplet curvature changes, it may not be possible for the interface to maintain condition II of the equilibrium at every location on the CL. The CL depins from those pillars where this condition can no longer be satisfied. The CL jumps and pins again at the next suitable location where both equilibrium conditions are satisfied. In a practical scenario, during the recede stage, the CL does not execute a jump until the disk has moved a certain distance. The droplet morphology changes every time the disk is moved to satisfy the two equilibrium conditions mentioned above. The droplet can attain equilibrium morphologies for the different positions of the disk since the CL is pinned at a sharp edge (of the pillars) and can remain in equilibrium for a range of contact angles, as dictated by the Gibbs inequality \cite{Gibbs1961}. The CL only starts to move when the local contact angle (at certain locations) reduces to a value equal to or less than $\theta_{\rm{e}}$ ($\theta_{\rm{R}}$ for real surfaces \cite{Gibbs1961}. This criterion for the CL depinning has also been used by others \cite{mognetti2010modeling}). The validity of the Gibbs criterion in CL depinning has been experimentally demonstrated by \citeauthor{paxson2013self} \cite{paxson2013self}.
In this work as well, the CL pinning at the edges of a pillar is governed by the Gibbs criterion \cite{Gibbs1961}, and the depinning starts when the local contact angle at any location on the pillar falls below 
$\theta_{\rm{e}}$ (or $\theta_{\rm{R}}$ for real surfaces with a non-zero inherent hysteresis) \cite{dyson1988contact,paxson2013self,mognetti2010modeling}. Different approaches have been proposed in the literature to model the dynamics of receding CL, notably the \textit{stick-slip} motion, where the CL sticks to the trailing (towards the air side) \cite{gao2006lotus} or leading (towards the liquid) \cite{dorrer2011micro} edge of the pillar before executing a jump. 
\citeauthor{yeong2015microscopic}  \cite{yeong2015microscopic} showed experimentally that the receding CL follows a \textit{slide-stick} motion where it travels a small distance on the tops of the pillar moving towards the leading edge before executing a jump. Because we can only calculate the equilibrium droplet morphologies, we cannot capture CL movement on the pillar tops.

When the CL depins from a pillar, a capillary wave is generated \cite{nguyen2016depinning} which eventually gets dampened due to the fluid's viscosity. For a typical drop adhesion measurement with a $2$ $\mu$L water drop in air, the Ohnesorge number, $Oh = \mu/\sqrt{\rho \sigma_{\rm{la}} L}\sim0.004$, where $\rho$ and $\mu$ are the density and viscosity of the liquid. This indicates that inertial and capillary effects dominate over viscosity, and that it is possible the capillary wave generated during the CL depinning may result in further depinning events at other locations if it has sufficient energy. Any such catastrophic instability will present itself in a deviation from the typical \textit{stick-slip} motion of the CL, which we did not observe in any of the experimental data used for model validation in the present study. In addition, it has not been observed in previous experimental measurements \cite{liimatainen2017mapping, daniel_mapping_2019, paxson2013self}. Notably, \citeauthor{nguyen2016depinning} \cite{nguyen2016depinning} experimentally investigated the generation of capillary waves during CL depinning in a sliding microliter droplet on a pillared superhydrophobic surface. Their observations indicate that the contact line retains its characteristic \textit{stick–slip} motion while the capillary wave propagates through the droplet. Therefore, for a microliter droplet moving at an infinitesimal speed in contact with a surface with micron sized pillars, the next suitable morphology of the droplet after a CL depinning event can be obtained by the two equilibrium conditions described earlier. A similar analysis, but in the context of microscale dynamics of the interface in Wenzel wetting case has been presented in a previous work \cite{kumar2024numerical}. 

Upon subsequent travel of the disk, when the droplet cannot remain in equilibrium while attached to the surface for a particular disk position, the droplet detaches from the surface (\textit{detach} stage).

\subsection{Numerical model}
\label{sec:numerical_model}

In this section, we present the numerical model for simulating droplet probe microscopy discussed in \S \ref{sec:physical_model}. We use the open-source software Surface Evolver (SE) \cite{brakke1992software} to implement the model. SE is a numerical optimization tool that works on the principle of gradient descent. To obtain equilibrium droplet morphologies, we use SE to minimize the total energy as given in Eq. (\ref{eqn:enegy3}). 
More details on the operation of SE can be found in the software manual \cite{brakke1994manual}.


\subsubsection{Computational domain}
Fig. \ref{fig:cv}b shows the computational domain used in the present study. Due to the inherent symmetry of the system, we simulate only an octant of the full droplet. A small droplet of a few microliters in volume is attached to a circular disk (a few millimeters in diameter) that can move in the $z$ direction with the disk. The bottom surface of the domain is superhydrophobic, composed of cylindrical pillars of diameter $d$ and height $h$ arranged in a square pattern with an inter-pillar spacing $d_{\rm{avg}}$. The droplet-disk contact is constrained by a level-set constraint ($x^2+y^2-R_{\rm{disk}}^2 = 0$) that keeps the droplet pinned to the edges of the circular disk (with radius $R_{\rm{disk}}$). Young's angle boundary condition is applied at the CL. The symmetry boundary condition is applied to the left wall ($y=0$) and the right wall ($y=x$) of the domain. Cylindrical pillars are generated using three-dimensional shapes known as superquadrics \cite{barr1981superquadrics} represented by the following equation:
\begin{equation}
    \left(\frac{x-x_0}{r} \right)^2 + \left(\frac{y-y_0}{r} \right)^2 + \left( \frac{z}{h} \right)^n = 1.
    \label{eqn:superquad}
\end{equation}
The parameter $n$ controls the sharpness of the edge and the coordinates ($x_0,y_0$) define the location of the pillars on the surface. The constants $r$ and $h$ are the radius and height of the pillars, respectively. In this work, we use $n=250$ to model pillars.

\subsubsection{Capturing interfacial dynamics}
\label{sec:interface_dynamics}

Here we discuss the numerical approach to simulate the advancing and receding of the CL during the \textit{advance} and \textit{recede} stages, respectively. The approach used here is similar to the incremental advance approach proposed by \citeauthor{kumar2024numerical} \cite{kumar2024numerical}, with the difference being that instead of a portion of the interface, the entire droplet is simulated.

\begin{figure*}
    \centering
    \includegraphics[width=0.65\linewidth]{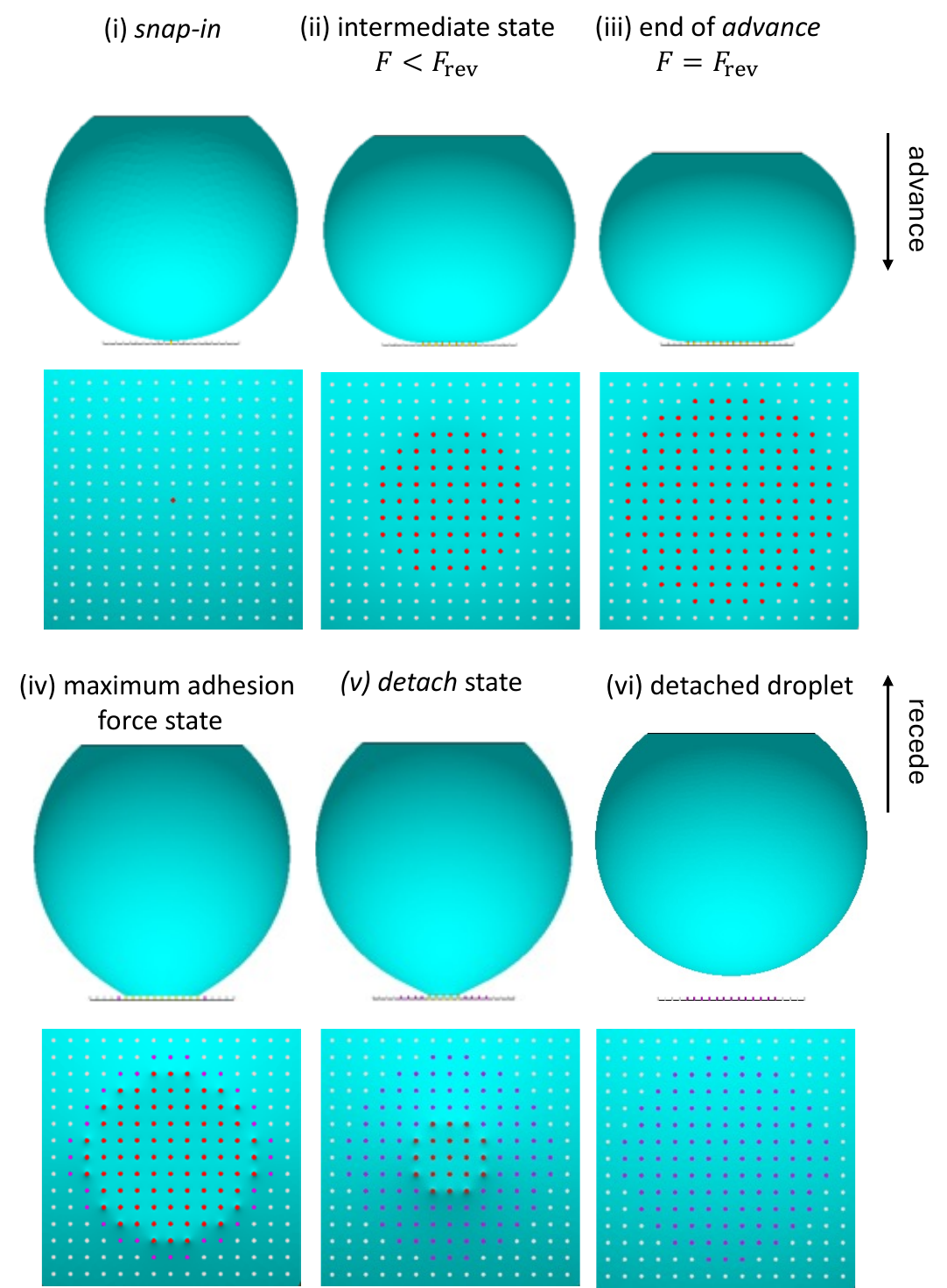}
    \caption{Front and bottom view of the equilibrium droplet morphologies during the \textit{advance} and \textit{recede} stages obtained via numerical simulation on a superhydrophobic surface with $\theta_{\rm{A}}=111.2$\textdegree, $\theta_{\rm{R}}=98.7$\tc and $\phi=0.05$ and a droplet of volume $V=1.6$ $\mu$L. In the bottom view of the droplet the red, magenta and white colors are used to indicate the pillar tops in contact with the liquid, pillar tops where the CL has depinned and the pillar tops which were not in contact with the liquid at any point of time during the entire process, respectively. The high curvature of the liquid-air interface lying between the pillars can be seen in the bottom view of the equilibrium morphologies during the \textit{recede} stage.
    }
    \label{fig:algo_combined}
\end{figure*}

Fig. \ref{fig:algo_combined} shows the simulation snapshots showing the \textit{snap-in} and \textit{advance} stages. The flow chart of the algorithm used to simulate these stages is shown in Figure S1 in the Supplemental Material \cite{SupplementalMaterial}. The simulation starts with a droplet of a given volume attached to the disk. The droplet is then gradually lowered until it contacts the pillar tops. The $z$ coordinate of the lowest facet of the droplet (average of the three vertices of the facet) is tracked to determine the precise instant of contact. When the droplet contacts the surface, the intersection between the liquid-air mesh and the pillar mesh is obtained. The intersecting facets (and edges and vertices) of the interface are then constrained to lie on the pillars using Eq. (\ref{eqn:superquad}). To account for the change in the total energy of the system, the interfacial tensions of the intersecting facets are changed from $\sigma_{\rm{la}}$ to $-\sigma_{\rm{la}}\cos\theta_{\rm{e}}$ (or $-\sigma_{\rm{la}}\cos\theta_{\rm{A}}$ for real surfaces). This change in the interfacial tension signifies the change in the type of interface from the liquid-air to the solid-liquid interface. An equilibrium morphology of the droplet is obtained by subsequent energy minimization and mesh refinements (see \S S7 in SI of \cite{kumar2024numerical}), which represents the droplet morphology at \textit{snap-in}. 
%

To advance the droplet, the disk is moved towards the surface by a  small amount ($\delta H$). With gradual minimization and mesh refinements, the equilibrium droplet morphology is obtained for the new position of the disk. To track the CL advancement, the regular gradient descent iteration in SE (`g') is replaced by a modified iteration step (`gadv'). In addition to the default gradient descent step, the `gadv' command also tracks the vertical position ($z$ coordinates) of the vertices of the liquid-air interface whose projection on the $x-y$ plane lies within the perimeter of the pillar tops. A suitable scale factor of the minimization iterations is also maintained through the simulation, demonstrating that the simulations are numerically stable (see Figure S4 in the Supplemental Material \cite{SupplementalMaterial}).  When the liquid-air interface approaches a pillar, `gadv' places the liquid-air facets (and edges and vertices) whose projection lies within the perimeter of the pillar top on the constraint of that pillar (Eq. (\ref{eqn:superquad})). The interfacial energies of the newly constrained liquid-air interface are also changed to $-\sigma_{\rm{la}}\cos\theta_{\rm{e}}$ (or $-\sigma_{\rm{la}}\cos\theta_{\rm{A}}$ for real surfaces) to accommodate the changes in total energy of the system. The energy minimization is performed to obtain the equilibrium droplet morphology corresponding to the new position of the disk, and at the same time the intersection of the liquid-air interface pillars is monitored to advance the CL. This process is repeated until the disk travels a pre-determined distance or the force exerted on the surface by the droplet reaches a certain pre-determined maximum value. We use the principle of virtual work using Lagrange Multipliers (see \cite{brakke1994manual} pp. 242-243) for calculating the adhesion force from equilibrium droplet morphologies (also see \S S1 in the Supplemental Material \cite{SupplementalMaterial}).

\label{sec:recede_stage}
At the end of the \textit{advance} stage, the droplet is in a compressed state between the disk and the surface. To recede the droplet, the disk is moved away from the surface in small steps ($\delta H$) followed by subsequent steps of  energy minimization and mesh refinement (see snapshots in Fig. \ref{fig:algo_combined}). The flow chart of the algorithm used to simulate these stages is shown in Figure S2 in the Supplemental Material \cite{SupplementalMaterial}. The key during the \textit{recede} stage is to predict the CL depinning. 
In the SE model, a modified iteration command `grec' is used instead of the regular gradient descent command (`g'). This command computes the local contact angle ($\theta_{\rm{local}}$) at every point around the perimeter of each pillar on the CL and initiates the CL depinning from a pillar if $\theta_{\rm{local}} \neq \theta_{\rm{e}}$ (or $\theta_{\rm{R}}$ for real surfaces) at any point on the CL on that particular pillar. The CL depinning in the model is achieved by removing the pillar profile constraint (Eq. (\ref{eqn:superquad})) from the pinned vertices (and facets and edges) and changing the interfacial tension of the pinned facets from $-\sigma_{\rm{la}}\cos\theta_{\rm{e}}$ (or $-\sigma_{\rm{la}}\cos\theta_{\rm{R}}$ for real surfaces) back to $\sigma_{\rm{la}}$. This ignores any micro-droplet generation during CL receding \cite{wong2020microdroplet}, which is an acceptable assumption since changes in total interfacial energy due to loss of a small droplet volume can be neglected and we do not expect any droplet of substantial volume to remain on the tops of the pillars due to a suitably high receding contact angle considered in the present work \cite{darvish2026water}. After the CL depins from a pillar, subsequent energy minimization and gradual mesh refinements were performed to find the equilibrium droplet morphology. It should be noted that the next possible equilibrium droplet morphology is governed by the two equilibrium conditions outlined in \S \ref{sec:physical_model} being satisfied. The modified `grec' command in the algorithm makes sure that the Young's angle boundary condition (condition II) is being satisfied at all the times while the entire droplet is moving towards a minimum energy morphology. If this condition is not met, that is, $\theta_{\rm{local}} \neq \theta_{\rm{e}}$ (or $\theta_{\rm{R}}$ for real surfaces), then the solid-liquid mesh on the pillar is changed back to the liquid-air mesh and the CL is depinned from that particular pillar. As the \textit{recede} stage progresses, there comes a point when an equilibrium morphology of the droplet cannot be obtained under the given disk position. At that instance the droplet detaches from the surface, indicating the \textit{detach} stage.

The model outlined above assumes zero inherent hysteresis, that is, the CAH exhibited by the superhydrophobic surface is due to the presence of microscopic pillars only. It assumes that the parent surface is free from hysteresis, and the inherent advancing ($\theta_{\rm{A}}$) and receding ($\theta_{\rm{R}}$) contact angles are the same, that is, $\theta_{\rm{A}}=\theta_{\rm{R}}=\theta_{\rm{e}}$. 
However, in reality, there is always a finite inherent hysteresis present in the system, i.e. $\theta_{\rm{A}}>\theta_{\rm{R}}$,
due to the presence of roughness at length scales below the primary roughness length scale (ie, below the pillar size and spacing). To consider the effect of inherent CAH in the numerical model, we take $\theta_{\rm{A}}$ and $\theta_{\rm{R}}$ instead of $\theta_{\rm{e}}$ during the \textit{advance} and \textit{recede} stages, respectively (see \cite{he2004contact} and \cite{kumar2024numerical} for a similar approach). However, the use of $\theta_{\rm{R}}$ instead of $\theta_{\rm{e}}$ during the \textit{recede} stage poses a numerical problem, especially if $\theta_{\rm{R}}<90$\textdegree. The error manifests itself as a wetting transition from the Cassie-Baxter to Wenzel state if the pillars are sparsely positioned and/or the inherent receding angle is too low. This wetting transition is nonphysical since the CL has to advance down the sides of the pillars, which will only happen if the local contact angle is greater than or equal to the inherent advancing angle \cite{papadopoulos2013superhydrophobicity}. To avoid this nonphysical behavior, either a one-sided constraint could be applied to the CL vertices (see \S 3.6 in the manual \cite{brakke1994manual}) or the vertices of the solid-liquid interface could be fixed (the approach used in this study) as shown in the algorithm flow chart in \S S2 in the Supplemental Material \cite{SupplementalMaterial}. This also modifies the condition II of the equilibrium, that is, Young's angle boundary condition. Now the droplet stays in equilibrium if $\theta_{\rm{R}}<\theta_{\rm{local}}<\theta_{\rm{A}}$ and the CL depinning is initiated only when $\theta_{\rm{local}}\leq\theta_{\rm{R}}$, instead of the condition $\theta_{\rm{local}} \neq \theta_{\rm{e}}$ which can only be used when the parent surface has zero inherent hysteresis.

The Surface Evolver code used to simulate the droplet probe microscopy has been made available at \cite{Kumar2026Code} for use and further development. To ensure that energy minimization works properly, it is imperative that throughout the simulation the scale factor is maintained at a suitable value of $\sim 0.2$ (see Supplemental Material S6 \cite{SupplementalMaterial}). A video has been provided in the Supplemental Material \cite{SupplementalMaterial} (Movie S3) that shows a typical droplet probe microscopy simulation.

\section{Results}
In this section, we present results for typical droplet probe microscopy simulated by the energy  minimization approach discussed in \S \ref{sec:interface_dynamics}. We discuss the variation in droplet and CL morphologies, and the forces that act between the droplet and a surface as the droplet contacts, advances, recedes, and then detaches from it. The surface considered is a superhydrophobic surface with microscopic cylindrical pillars with an aspect ratio ((pillar height divided by the diameter) of 2.0 and inherent advancing and receding contact angles of $\theta_{\rm{A}}=111.2$\tc and $\theta_{\rm{R}}=98.7$\tc, respectively. 
The simulations were performed in non-dimensional form using a characteristic length scale $l_0=10$ $\mu$m and a characteristic surface tension of $\sigma_0=72.8$ mN/m. A droplet of volume, $V=1.6$ $\mu$L was used to generate the results. In non-dimensional form the pillars have a diameter of 1, height of 2, and the droplet has a volume of $1.6 \times 10^6$. 

A complementary experimental study was carried out to validate the model. Superhydrophobic surfaces composed of cylindrical pillars (20 $\mu$m high and 10 $\mu$m in diameter) arranged in a square array were used for the measurements. These surfaces were etched onto a silicon wafer using a dry etching technique and were subsequently coated with thin layers of chromium (5 nm) and gold (25 nm) by physical vapor deposition \cite{kumar2024experimental}. Two different pillar area fractions ($\phi=0.05$ and $\phi=0.30$) were used in this study. These were fabricated by varying the inter-pillar spacing while keeping the pillar diameter constant. The surfaces were hydrophobized before measurement by depositing a monolayer of octadecanethiol from a 1 mM ethanol solution (\cite{kumar2024experimental}). The contact angle measured on the flat octadecanethiol coated surfaces exhibited advancing and receding angles of 111.2\tc and 98.7\textdegree, respectively. We used a DataPhysics DCAT-21 tensiometer to measure the adhesion force between a 1.6 $\mu$L ultra pure DI water droplet and the surface. In the moving disk mode, the stage was moved at a speed of 25 $\mu$m/s during the \textit{recede} stage. Detailed experimental procedure is given in Appendix \ref{SI:experiments}. 

\subsection{Droplet and contact line morphology}
Fig. \ref{fig:algo_combined} shows the equilibrium morphologies of a droplet during a typical droplet probe microscopy simulation. 
The footprint of the droplet is also shown in the figure capturing different stages of the simulation. We observe that the CL is non-continuous but discrete, lying on individual pillars that enclose the solid-liquid interfacial area, which is shown in red in the figure.
Therefore, the conventional definition of the CL as the continuous line bounding the solid-liquid interface cannot be used for the Cassie-Baxter wetting state. To aid discussion, we adopt the concept of an apparent CL as previously described in the literature, defined as the intersection of the liquid-air interface with a plane parallel to the general surface and a certain distance from the tops of the pillars ($\Delta z$). This definition is similar to the definition adopted by \citeauthor{dorrer2007contact} \cite{dorrer2007contact}.

\begin{figure*}
    \centering
    \includegraphics[width=\linewidth]{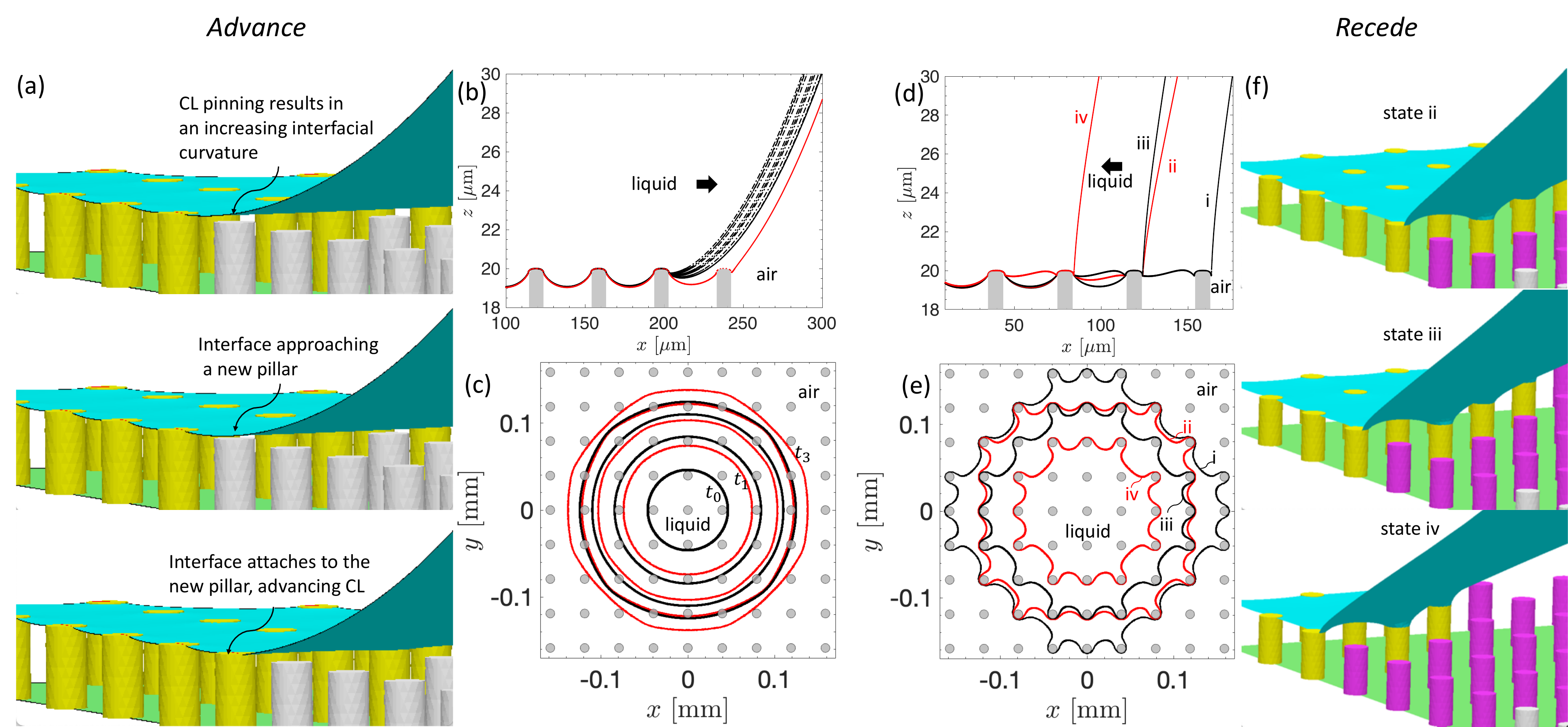}
    \caption{Depiction of the dynamics of the interface during the \textit{advance} and \textit{recede} stages respectively for a 1.6 $\mu$L droplet on a superhydrophobic surface with $\theta_{\rm{A}}=111.2$\textdegree, $\theta_{\rm{R}}=98.7$\tc and $\phi=0.05$. (a) Equilibrium interfacial morphologies showing how the interface advances. The CL stays pinned on a set of pillars while the liquid-air interface curves forward until it contacts the next set of pillars, thereby advancing the CL. The equilibrium interfacial morphologies on the $x-z$ plane is shown in (b). The pinned states are shown by dashed-dotted lines while the equilibrium states just before and  after the advancement of the CL are shown by the solid black and solid red lines respectively. (c) Shows the apparent CL measured at a distance of 1 $\mu$m from the pillar tops. The CL starts as a circle ($t_0$) and eventually acquires a shape resembling that of an octagon as it advances (eg. $t_1 \to t_3$. ). During the \textit{recede} stage the CL moves in its characteristic \textit{stick-slip} motion, executing jumps between the first and the second critical states shown by solid back and red lines, respectively, in (d) which shows the equilibrium interfacial morphologies on the $x-z$ plane. The corresponding CL morphologies (at $\Delta z=1$ $\mu$m) are shown in (e). (f) Equilibrium interfacial morphologies corresponding to the critical states ii, iii and iv in (d). The yellow colored pillars indicated pinned pillars, magenta represents the depinned pillars, and the white color represents the pillars which were never contacted by the liquid.}
    \label{fig:adv_morpho}
\end{figure*}

Fig. \ref{fig:adv_morpho}a shows the equilibrium interfacial morphologies that capture the microscale dynamics of the interface during a typical \textit{advance} stage. We observe that the liquid-air interface curves until it touches a new pillar, at which instant the CL advances \cite{schellenberger2016water}. Fig. \ref{fig:adv_morpho}b shows the liquid-air interface on the $x-z$ plane capturing the advancing motion. The dotted lines represent the interfacial morphology when the CL is pinned onto a set of pillars, and the solid line represents the equilibrium morphology just after the CL advancement. Due to the particular square arrangement of the pillars on the surface, the distance between the pillars along the $\psi=0$\tc direction ($d_{\rm{avg}}$) is smaller than that along the $\psi=45$\tc direction ($\sqrt{2}d_{\rm{avg}}$). Here, the angle $\psi$ defines the direction of interfacial motion relative to the $x$ axis. Therefore, the CL advances preferably in the $\psi=45$\tc direction, resulting in a non-circular CL shape. Fig. \ref{fig:adv_morpho}c shows the evolution of the apparent CL ($\Delta z=1$ $\mu$m) during the \textit{advance} stage. The first (equilibrium state just before the interface touches the next pillar(s) in the advancing direction) and second (equilibrium state when the interface has just contacted a new pillar(s) in the advancing direction) critical states are shown as black and red curves, respectively. The CL starts as a circle ($t_0$) and slowly transforms into a shape that resembles an octagon as it advances ($t_3$). During the \textit{advance} stage, the microscale deformations in the liquid-air interface are contained in a very narrow region around the actual CL. Therefore, we only see distortion of the apparent CL when measured close to the pillars ($\Delta z=1$ $\mu$m). In Fig. \ref{fig:SI_CL}a in Appendix \ref{SI:adv_CL}, we show that the apparent CL preserves a circular shape when measured sufficiently away from the pillars ($\Delta z=20$ $\mu$m).  

The dynamics of CL during the \textit{recede} stage, unlike the rolling motion observed in the \textit{advance} stage, can be described by a \textit{stick-slip} motion \cite{yeong2015microscopic,paxson2013self,schellenberger2016water}. The CL moves by executing occasional jumps between the first and second critical states, for example $i \to ii$ and $iii\to iv$ in Fig. \ref{fig:adv_morpho}d. Most of the time, the CL is pinned onto the pillars, while the interface continues to move macroscopically. This introduces large deformations in the liquid-air interface in a microscopic region close to the CL (see Fig. \ref{fig:adv_morpho}f). These interfacial deformations result in a highly distorted CL during the \textit{recede} stage. Fig. \ref{fig:adv_morpho}e shows the apparent CL ($\Delta z=1$ $\mu$m) corresponding to the first ($i$ and $iii$) and the second ($ii$ and $iv$) critical states shown in black and red, respectively, in Fig. \ref{fig:adv_morpho}d. The receding CL morphology is more complex compared to that during the advancing motion due to the highly distorted liquid-air interface near the CL. However, as the distance from the pillars is increased, the interfacial deformations decrease and the apparent CL takes a less distorted shape. This is shown in Fig. \ref{fig:SI_CL}b in Appendix \ref{SI:adv_CL}, where the apparent CL at $\Delta z=20$ $\mu$m is shown during the \textit{recede} stage on a surface with $\phi=0.05$. Interestingly, when sufficiently away from the pillars, such as $\Delta z=20$ $\mu$m as shown in Fig. \ref{fig:SI_CL}b, the apparent receding CL starts from a shape that resembles an octagon and ends up appearing as a circle just before detachment. This is opposite to the evolution of the apparent CL during \textit{advance} stage from a circle to that resembling an octagon. This evolution of the CL shape is due to the anisotropic nature of energy dissipation during interfacial dynamics \cite{kumar2024numerical}. The details of the CL dynamics, as well as the associated dissipation of energy and its relationship to contact angle hysteresis can be studied using the presented model, but are out of the scope of this study.



\subsection{Adhesion force}
\label{sec:force}

In this section, we discuss the variation in the net vertical force ($F$) between the droplet and the surface during a typical droplet probe microscopy (see \S S1 in the Supplemental Material \cite{SupplementalMaterial} for the numerical scheme for the calculation of force). 
\begin{figure*}
    \centering
    \includegraphics[width=0.50\linewidth]{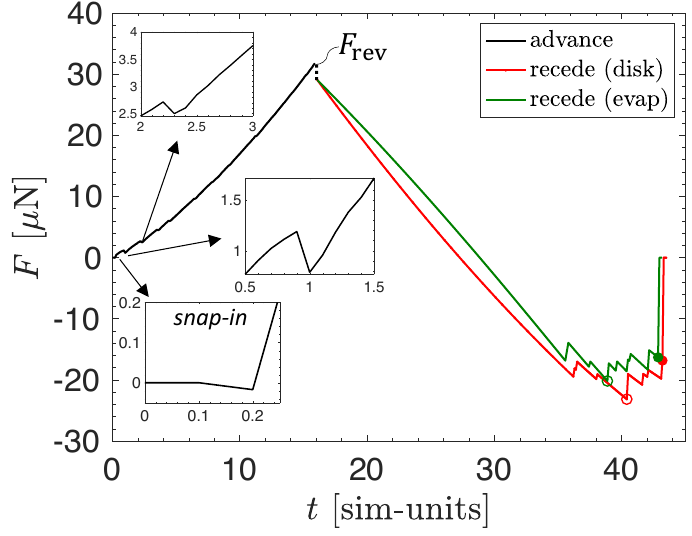} 
    \caption{Variation in the net vertical force $F$ with time $t$ (in simulation units), between a droplet and a superhydrophobic surface during a typical droplet probe microscopy simulation with maximum and detachment forces shown by empty and filled circles respectively. The force variation for the \textit{recede} stage is shown for both the moving disk (in red) and evaporation (in green) modes. The simulation time for the evaporation mode is scaled down to fit in the same plot. Inset shows the sharp drop in the force during \textit{snap-in} and two CL advancement moments. 
    The maximum and detachment forces obtained by each approach are comparable. 
    The force curves were generated for a 1.6 $\mu$L droplet on a surface with $\theta_{\rm{A}}=111.2$\textdegree, $\theta_{\rm{R}}=98.7$\tc and $\phi=0.05$.}
    \label{fig:force_curve}
\end{figure*}
Fig. \ref{fig:force_curve}a shows the variation in $F$ with time in simulation units (assuming a unit disk velocity). Note that the numerical simulations only calculate the consecutive equilibrium stages of the droplet that is moving at a much slower velocity compared to the CL velocity during jumps between the first and second critical states (discussed in \S III A). The simulation time is based on the displacement of the disk and is assumed to be moving at a unit velocity. One alternative is to plot the force variation with the displacement of the disk. However, we choose to show the force variation with time to keep it consistent with the current convention in displaying the results of the droplet probe microscopy  \cite{daniel_probing_2023,liimatainen2017mapping}. The choice of variable (time vs disk displacement) does not change the nature of force variation, as shown in Fig. \ref{fig:F_vs_z} in Appendix \ref{SI:F_vs_z}; we chose to plot against time as it is relatively easier to understand.

A small drop in $F$ is observed when the droplet contacts the surface, that is, the \textit{snap-in} stage. This is due to the sudden drop in energy as the rapidly moving CL dissipates energy \cite{Harvie_2024,jiang2019generalized,patankar2010hysteresis,liu2015dynamic,smith2018rolling}.
As the CL advances (the disk moves downward), $F$ appears to increase linearly. However, a closer look reveals a sawtooth variation in the force during the \textit{advance} stage. Every time the interface contacts a new set of pillars and the CL advances, there is a drop in total energy of the system due to dissipation (see Fig. \ref{fig:dissipation} in Appendix \ref{SI:dissipation}) that results in a small but sharp drop in $F$, as shown in the insets. A similar sawtooth variation in force during the advancement of CL was observed experimentally by \citeauthor{liimatainen2017mapping} \cite{liimatainen2017mapping}. 
The \textit{advance} stage continues until the forces reach a certain maximum value, herein termed the reversal force ($F_{\rm{rev}}$). At this stage, the direction of motion of the disk is reversed and the droplet is pulled away from the surface. It should be noted that the sharp drop in the force observed at the reversal point (shown by dashed lines in the figure) is due to the change in the Young's angle boundary condition at the CL from $\theta_{\rm{A}}$ to $\theta_{\rm{R}}$. 

During the \textit{recede} stage, the force magnitude increases (in the opposite direction). The receding motion of the CL is characterized by its typical \textit{stick-slip} movement that is reflected in a sawtooth variation in the $F$. The force increases linearly during CL pinning, drops sharply when the CL jumps due to the dissipation of energy (see Fig. \ref{fig:dissipation} in Appendix \ref{SI:dissipation}), and then increases again linearly when the CL pins again at a new location. The force continues to increase until it reaches a maximum value $F_{\rm{max}}$, followed by the detachment of the droplet at a relatively lower force $F_{\rm{detach}}$. The difference between the distinctive peaks and troughs of the sawtooth variation during the \textit{recede} stage is considerably greater than that during the \textit{advance} stage, because of the higher dissipation of energy during the receding motion as compared to the advancing motion on a superhydrophobic surface.

\subsection{Droplet evaporation}
\label{sec:evap}

In the above discussion, we showed how the force between a microliter droplet and a pillared superhydrophobic surface varies over time during typical droplet probe microscopy. Up to this point we have assumed that the rate of evaporation of the liquid is negligible and that the droplet volume remains constant during the entire process. However, in reality, if the drop motion is slow enough, then evaporation will cause a change in the droplet volume which may lead to a possible change in the CL dynamics, and subsequently the force. Here, we discuss an alternative approach to measuring the adhesion force between a droplet and the surface that relies on droplet evaporation. The droplet is not pulled away from the surface: instead the disk is kept stationary for the entire duration of \textit{the recede} stage and the droplet is allowed to evaporate, as done for example by \citeauthor{liimatainen2017mapping} \cite{liimatainen2017mapping}. Due to evaporation, the principal radii of curvature of the droplet decrease, resulting in an increase in Laplace pressure (see Figure S4 in the Supplemental Material \cite{SupplementalMaterial}). As the size of the droplet reduces, the CL depins from the pillars and jumps to a new equilibrium position, similar to what we observed when the droplet was moved away from the surface. The variation in $F$ during the \textit{recede} stage is shown in Fig. \ref{fig:force_curve} in green color. Note that the \textit{recede} stage in the evaporation mode takes much longer compared to the moving disk mode. However, we scaled the simulation time for the evaporation mode to fit both the evaporation and moving disk data in the same plot. For the evaporation mode as well, $F$ varies in a sawtooth manner, reaching a maximum ($F_{\rm{max}}$) before the droplet detaches from the surface with a lower force $F_{\rm{detach}}$. 
Compared with the moving disk mode, we observe that both approaches give similar force variations, with $F_{\rm{max}}$ differing only by a small amount ($\approx3$ $\mu$N). Therefore, numerically both approaches are equivalent and, either way, by moving the disk or by letting the droplet evaporate during the \textit{recede} stage, we will get a similar variation in the net vertical force. 
Finally, we end this section by commenting that in a typical force tensiometer setup, the droplet weight at the start of the experiment is subtracted, and the tensiometer only measures the interaction force between the droplet and the surface. However, as the droplet evaporates, the change in weight as a result of the loss of mass is not accounted for by the tensiometer, which results in different force curves for evaporation and moving disk modes. 
This is discussed in the next section, where we compare the numerical findings with experimental results.

\subsection{Comparison with experiments}
\label{sec:exp_comp}
\begin{figure*}
    \centering
    \includegraphics[width=0.85\linewidth]{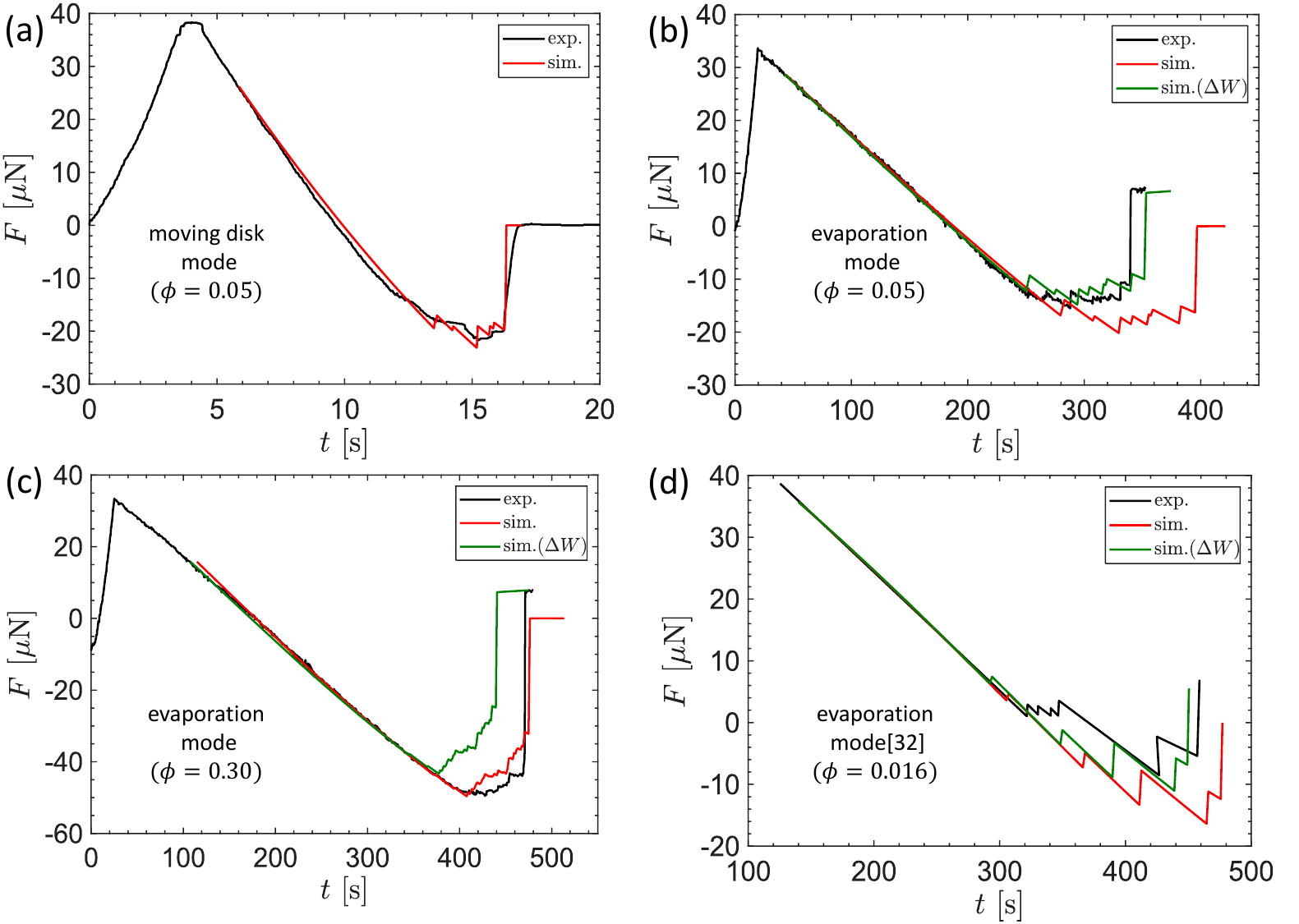} 
    \caption{Variation in the net vertical force ($F$, in micronewtons) with time ($t$, in seconds) measured experimentally and calculated via energy minimization simulations. For both the moving disk and evaporation modes, simulation time is scaled linearly to match the initial slope of the experimental force curve during the \textit{recede} stage. \textit{Advance}, \textit{recede} and \textit{detach} stages are shown in (a), (b) and (c) with $t=0$ representing the moment the droplet contacts the surface. $F$ vs $t$ for a droplet of volume $V=1.6$ $\mu$L and a superhydrophobic surface with $\theta_{\rm{A}}=111.2$\textdegree, $\theta_{\rm{R}}=98.7$\tc and pillar area fractions $\phi=0.05$ and $\phi=0.30$ are shown in (b) and (c) respectively.
    The force variation with an additional term for the loss in droplet weight due to evaporation (i.e. $F+\Delta W$) is also plotted in green in (b) and (c). 
    (d) $F$ vs $t$ variation from \citeauthor{liimatainen2017mapping} \cite{liimatainen2017mapping} (in black) and simulation results for the evaporation mode considering change in droplet weight (in green) and without considering change in droplet weight (in red) due to evaporation. The droplet has an initial volume of $V=1.5$ $\mu$L and the surface has $\theta_{\rm{A}}=111.0$\textdegree, $\theta_{\rm{R}}=85.1$\tc and $\phi=0.016$.}
    \label{fig:exp_vs_sim}
\end{figure*}
In \S \ref{sec:force} we discussed the variation in the adhesion force between a droplet and a superhydrophobic surface during a typical droplet probe microscopy. The force curve has a distinct sawtooth variation characterized by the maximum force ($F_{\rm{max}}$) and the detachment force ($F_{\rm{detach}}$) respectively. Here, we compare the numerical results with the experimental measurements of the adhesion force. In Fig. \ref{fig:exp_vs_sim}a we show the variation in $F$ with time (s) experimentally measured between a droplet 1.6 $\mu$L in volume and a superhydrophobic surface with $\phi=0.05$, $\theta_{\rm{A}}=111.2$\tc and $\theta_{\rm{R}}=98.7$\textdegree. We also plot the force variation obtained by numerical simulations. Since the numerical simulations only calculate the equilibrium interfacial morphologies, a pseudo time is obtained by assuming a certain velocity for the movement of the disk during the \textit{recede} stage such that the initial slope of both the curves at the start of the \textit{recede} stage is matched. 
We observe a good agreement between the experimental and simulation results, giving confidence that the numerical model is capable of accurately predicting the maximum adhesion force ($F_{\rm{max}}$), which is a characteristic of the surface. 
An important point to note here is that a typical commercial force tensiometer is not capable of capturing the peaks and troughs of the force variation during the \textit{recede} stage because of limited sensitivity. This would require a much more sensitive tensiometer setup such as the one recently developed by \citeauthor{liimatainen2017mapping} \cite{liimatainen2017mapping}. However, the numerical model presented here is capable of resolving the small force variations and can complement the results obtained using any commercial force tensiometer. 

Now we compare the numerical results against the experimental data for the case of an evaporating droplet. In Fig. \ref{fig:exp_vs_sim}b we show the variation in $F$ with time measured experimentally (in black) for an evaporating droplet of an initial volume, $V=1.6$ $\mu$L placed on a superhydrophobic surface with $\phi=0.05$, $\theta_{\rm{A}}=111.2$\tc and $\theta_{\rm{R}}=98.7$\textdegree. The droplet is kept stationary during the entire \textit{recede} stage (experimentally, this was achieved by keeping the stage on which the surface was placed stationary). The first difference between the moving disk (or stage in the experiments) and evaporating droplet modes is the duration of the \textit{recede} stage, which is significantly greater for the latter. For example, for an area fraction of $\phi=0.05$ the \textit{recede} stage lasts approximately 300 s during the evaporation mode compared to approximately 10 s during the moving disk mode. This time duration increases even more as the pillar area fraction increases (see Fig. \ref{fig:exp_vs_sim}c). Another important difference is that, during  the evaporation mode, the tensiometer captures some of the peaks and troughs of sawtooth variation in $F$ with time during the \textit{recede} stage. This is due to a much slower rate of CL motion during the \textit{recede} stage in the evaporation mode. It may appear that by letting the droplet evaporate, a commercial tensiometer can be used to capture the variation in the adhesion force between the surface and the droplet. However, careful observation reveals that loss of droplet mass as a result of evaporation results in a slight reduction in the measured adhesion force, as seen by a non-zero residual force in Figs. \ref{fig:exp_vs_sim}b and c after the droplet has detached from the surface. 
The magnitude of this residual force ($\approx$ 7.0 $\mu$N in the particular case shown in Fig. \ref{fig:exp_vs_sim}b) is similar to the change in the weight of the droplet due to evaporation. As an example, in Fig. \ref{fig:ras_area}a in Appendix \ref{SI:time_scaling} we show the variation in $F$ with the change in droplet volume ($\Delta V$) obtained numerically ($\phi=0.05$, $\theta_{\rm{A}}=111.2$\tc and $\theta_{\rm{R}}=98.7$\textdegree). We observe that droplet detachment occurs at $\Delta V=0.64$ $\mu$L, which is equivalent to a weight of $6.26$ $\mu$N. The similar magnitudes of the change in droplet weight due to evaporation and the residual force after detachment during the experiments (evaporation mode) suggests that the unaccounted droplet weight may be causing the difference between the experimental and simulation results. To address this, we adjusted the numerically calculated adhesion force by accounting for the equivalent weight of the evaporated droplet mass $\Delta W$ at each equilibrium position, shown in green in Fig. \ref{fig:exp_vs_sim}b (see Appendix \ref{SI:time_scaling} for details regarding mapping of simulation data onto the experimental force curve). We observe that the adjusted force ($F+\Delta W$) obtained numerically is in good agreement with the experimental data, but not for every case (Fig. \ref{fig:exp_vs_sim}c). This may be because the volume of the droplet at the start of the \textit{recede} stage is unknown. The numerical simulations assume that the \textit{advance} stage is fast enough so that the loss of vapor as a result of evaporation is negligible. However, there may still be some loss of volume and the droplet may have a different volume at the start of the \textit{recede} stage compared to the volume at the start of the experiment (or simulation). It should be noted that droplet volume at the start of the \textit{recede} stage may be different from the initial volume due to the droplet evaporation during both the \textit{approach} and \textit{advance} stages. In Figs. \ref{fig:exp_vs_sim}b and c, even though the \textit{advance} stage for both the cases lasts for a similar duration, the \textit{approach} duration, that is, time duration before the droplet contacts the surface for $\phi=0.30$ was comparatively greater than that for the $\phi=0.05$ case (see Movie S5 in the Supplemental Material \cite{SupplementalMaterial}), which may have caused a high difference in the droplet volume at the start of the \textit{recede} stage for $\phi=0.30$. Finally, we compare the numerical model with the experimental data of \citeauthor{liimatainen2017mapping} \cite{liimatainen2017mapping}. We numerically calculated the adhesion force between a 1.5 $\mu$L droplet (initial volume) and a superhydrophobic surface with $\phi=0.016$, $\theta_{\rm{A}}=111.0$\tc and $\theta_{\rm{R}}=85.1$\textdegree. The simulations were performed in the evaporation mode. In Fig. \ref{fig:exp_vs_sim}d we show the experimental results in black and simulation data in red and green (considering the loss in droplet mass). We observe that the numerical model overpredicts the force, however there is good agreement between model and experiment when taking the loss in droplet mass ($F+\Delta W$)  into account as the measurement proceeds. The above results are encouraging and builds confidence in the model. Based on the above discussion we suggest that, by incorporating the loss in volume due to evaporation and using a correct droplet volume at the start of the \textit{recede} stage, the present numerical model may be used to predict the adhesion force acting on an evaporating droplet.
Apart from the change in the droplet weight due to evaporation, there are other factors that may need to be considered. For example, the evaporating droplet can result in a non-uniform change in the surface temperature of the droplet, generating surface tension gradients \cite{dekker2025hidden} affecting the variation of the force. In addition, vapor accumulation near the CL can affect the CL pinning by exerting a localized pressure force, resulting in a different variation in the force as the droplet evaporates. Because these effects were not incorporated into the numerical model, it is difficult to comment how much these factors would impact the overall droplet adhesion. A complete understanding of the effect of droplet evaporation requires a series of experiments under controlled evaporation, which is outside the scope of this work and will be considered in the future.

\section{Conclusion}
\label{sec:conclusion}

We have presented a quasi-static numerical method to simulate typical droplet probe microscopy on superhydrophobic surfaces composed of periodic pillar structures. The present model can predict the variation in the adhesion force as the different stages of droplet probe microscopy, namely, \textit{approach, snap-in, advance, recede}, and \textit{detach}, are performed. The model assumes: (1) the droplet remains in the Cassie-Baxter wetting state, (2) the interface is in equilibrium in contact with the surface, (3) constant and uniform surface tension coefficient, (4) negligible gravitational effects, and (5) the surface is chemically homogeneous with constant inherent advancing andreceding contact angles. As the model is based on quasi-static interfacial dynamics, it does not predict the rapid CL motion during the jumping event. However, for most cases where a microliter droplet is in contact with a pillared surface with a hydrophobic chemical coating, these assumptions are valid and therefore the model can be used for a range of droplet size, pillar geometry, inter-pillar spacing, and surface chemistry (Young's angle). 

Based on the numerical model, we simulate droplet probe microscopy with a microliter droplet ($V=1.6$ $\mu$L) and a superhydrophobic surface ($\theta_{\rm{A}}=111.2$\textdegree, $\theta_{\rm{R}}=98.7$\textdegree) with cylindrical pillars (height 20 $\mu$m and diameter of $10$ $\mu$m). We observe that the adhesion force between the droplet and the surface varies in a sawtooth manner as the droplet contacts the surface, is compressed between the stationary surface and the moving disk, and then is pulled away from the surface by moving the disk in the opposite direction. We observe that the peaks and troughs in this sawtooth variation occur at the moments when the contact line executes a jump and comes to an equilibrium state after executing a jump, respectively. During the \textit{recede} stage, the adhesion force reaches a maximum before the droplet detaches from the surface at a much lower force magnitude. We simulated a pure evaporation in which both the surface and the disk are not moved during the \textit{recede} stage, and the droplet is allowed to evaporate, triggering the detachment. We observe that the variation in the force during the \textit{recede} stage is similar to the case when the disk is moved and the evaporation is not considered. We compared our model with the experiments and observed a good agreement for the non-evaporating case. For the evaporating droplet, we observe that the model over-predicts the forces during the \textit{recede} stage, which is primarily due to the loss in droplet mass as it evaporates. Because only the initial weight of the droplet is adjusted in the microbalance reading, the difference in weight owing to evaporation remains unaccounted for in the experimental measurements. In addition to this, the presence of surface tension gradients as a result of changes in temperature upon evaporation and the localized vapor accumulation near the contact line may also affect the experimental measurements. We also studied the contact line dynamics during the \textit{advance} and \textit{recede} stages and observed that the apparent contact line evolves from a circular shape to that resembling an octagon during the \textit{advance} stage and then goes back to its original circular shape during the \textit{recede} stage. The reason for this interaction of the contact line is the anisotropy in the dissipation of energy during the contact line dynamics, which is left for future work. 

This work will help advance direct force measurement technology and has potential for many future studies. The present model can be used to study the stability of the Cassie-Baxter wetting state and the onset of the wetting transition to the Wenzel state. In this work, only a few pillar area fractions and a single set of inherent advancing and receding contact angles were considered. It would be interesting to see the effect of a variation in the area fraction of the pillar, the inherent advancing and receding contact angles, and the shape and aspect ratio of the pillars. As an application of the present model, we have presented a methodology for the quantitative characterization of superhydrophobic surfaces using droplet probe microscopy in \cite{kumar2025adhesion}.

\begin{acknowledgments}
J.D.B. received support from an Australian Research
Council Future Fellowship (FT220100319) funded by the Australian Government. The authors would like to thank Professor Marta Krasowska from the Adelaide University for her generous support during the experimental part of this study. 
\end{acknowledgments}
\section*{DATA AVAILABILITY}
The data used to generate the figures in this article are openly available at \cite{Kumar2026Figshare}. The Surface Evolver code used to simulate droplet adhesion is publicly available at \cite{Kumar2026Code}.

\appendix
\section{Evolution of the apparent CL at $\Delta z=20$ $\mu$m}
\label{SI:adv_CL}
\begin{figure}[h]
    \centering
    \includegraphics[width=\linewidth]{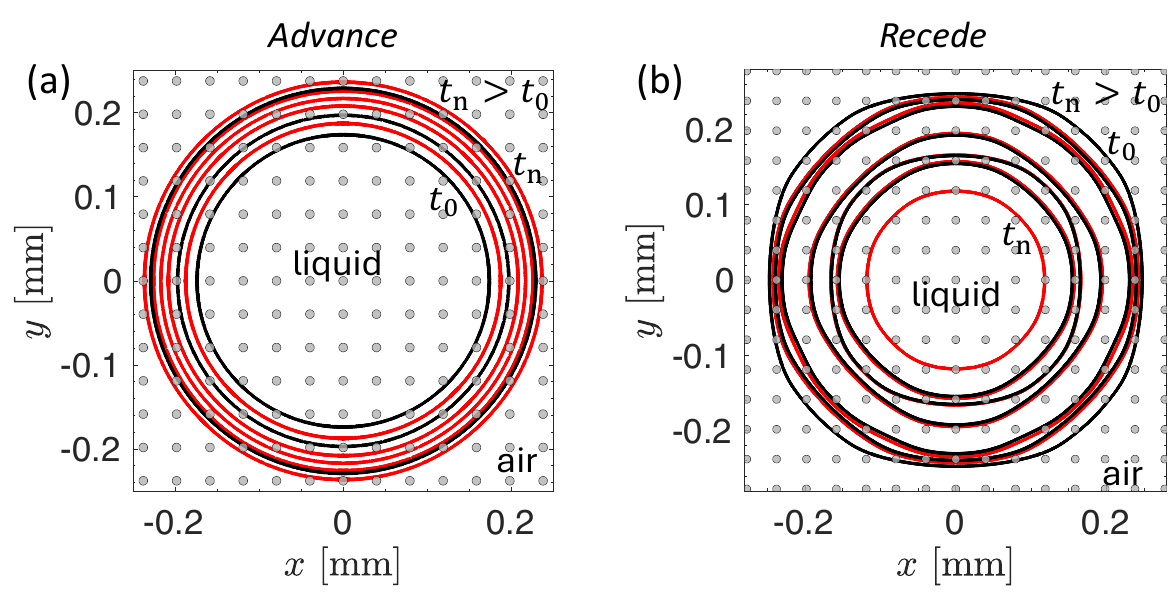} 
    \caption{Evolution of the apparent CL ($\Delta z=20$ $\mu$m) of a droplet of volume ($V=1.6$ $\mu$L) on a superhydrophobic surface with $\theta_{\rm{A}}=111.2$\textdegree, $\theta_{\rm{R}}=98.7$\tc and $\phi=0.05$ during the (a) \textit{advance} and (b) \textit{recede} stages respectively.}
    \label{fig:SI_CL}
\end{figure}
Figs. \ref{fig:SI_CL}a and b show the evolution of the apparent CL measured at $\Delta z=20$ $\mu$m from the tops of the pillars during the \textit{advance} and \textit{recede} stages, respectively. During the \textit{advance} stage, the apparent CL maintains a circular shape as the interface moves forward. However, during the \textit{recede} stage, the apparent CL evolves from a shape resembling that of an octagon to a shape resembling a circle. In fact, the apparent CL also evolves from a circular shape to a shape resembling an octagon during the \textit{advance} stage but is contained in a zone very close to the pillars (see Fig. \ref{fig:adv_morpho}c where the apparent CL at $\Delta z=1$ $\mu$m shows this evolution during the \textit{advance} stage).

\section{Variation in the droplet surface interaction force with the position of the disk}
\label{SI:F_vs_z}
\begin{figure}
    \centering
    \includegraphics[width=\linewidth]{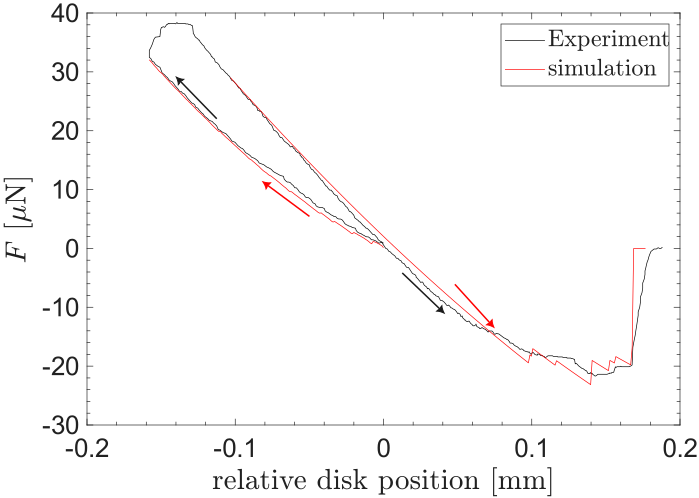} 
    \caption{Variation in the adhesion force ($F$) in $\mu$N between a small droplet and a superhydrophobic surface with the relative disk position during a typical droplet probe microscopy. Both the numerical (red) and experimental (black) results are shown. The numerical data for the recede stage has been plotted by uniformly shifting all the data points along the x-axis to account for the time taken by the equipment's stage to reverse its direction during the experiment.
    A droplet of volume $V=1.6$ $\mu$L and a superhydrophobic surface with $\theta_{\rm{A}}=111.2$\textdegree, $\theta_{\rm{R}}=98.7$\tc and $\phi=0.05$ were used in the simulation and the experiment.}
    \label{fig:F_vs_z}
\end{figure}

Fig. \ref{fig:F_vs_z} shows the variation in force $F$ between a droplet of volume 1.6 $\mu$L and a superhydrophobic surface ($\phi=0.05$, $\theta_{\rm{A}}=111.2$\tc and $\theta_{\rm{R}}=98.7$\textdegree) with the relative position of the disk. The relative position is obtained relative to the disk position when the droplet just contacts the surface (\textit{snap-in}). Good agreement is observed between the numerical and experimental data.

\section{Variation in total energy of the droplet}
\label{SI:dissipation}
\begin{figure}
    \centering
    \includegraphics[width=0.65\linewidth]{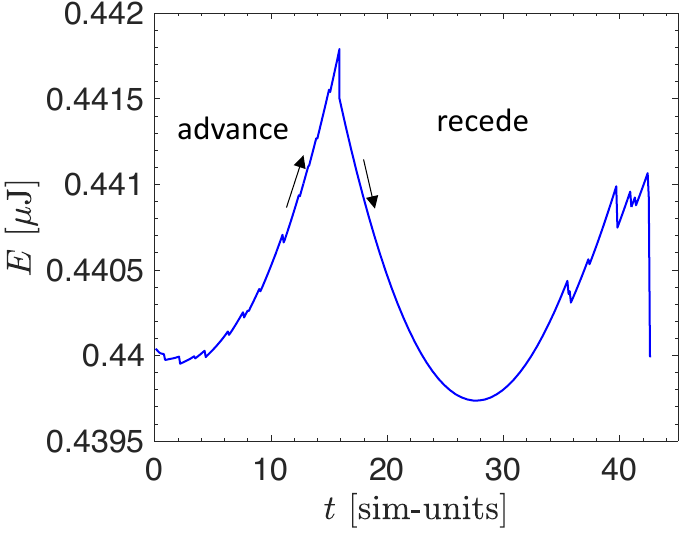} 
    \caption{Variation in total energy in $\mu$J (Eq. (\ref{eqn:enegy3})) with time in simulation units during the \textit{advance} and \textit{recede} stage of droplet probe microscopy obtained via energy minimization simulations. A droplet of volume $V=1.6$ $\mu$L and a superhydrophobic surface with $\theta_{\rm{A}}=111.2$\textdegree, $\theta_{\rm{R}}=98.7$\tc and $\phi=0.05$ were used in the simulation.}
    \label{fig:dissipation}
\end{figure}

Fig. \ref{fig:dissipation} shows the variation in the total energy of the droplet in $\mu$J (Eq. (\ref{eqn:enegy3})) during the \textit{advance} and \textit{recede} stages (moving disk mode) with time in simulation units. The total energy changes as the liquid-air and solid-liquid areas change during the coarse of the simulation run. A sudden drop in energy can be observed at locations where the CL moves at capillary velocities, dissipating energy. The energy dissipation also appears as a drop in $F$, shown in Fig. \ref{fig:force_curve} in the main text.  

\section{Mapping simulation data onto experimental force curve for the evaporation mode}
\label{SI:time_scaling}

To simulate the evaporation mode in Surface Evolver, we calculated the equilibrium droplet morphology for an initial drop volume $V_0$, and then reduce the droplet volume by a fixed percentage to calculate the next equilibrium shape, i.e. $\Delta V = -\alpha V$. Assuming that the droplet density is constant and there is a constant concentration gradient driving mass transfer, then the rate of change of the droplet volume is proportional to the instantaneous liquid-air interfacial area ($A_{\rm{la}}$), that is,
\begin{equation}
 \frac {dV}{dt} = -\beta A_{\rm{la}},   
 \label{eqn:time_scaling1}
\end{equation}
where $\beta$ is a constant depending on the gradient of vapor concentration, the diffusion coefficient and the droplet density. From the simulation results (Fig. \ref{fig:ras_area}), we observe that the interfacial area scales linearly with the droplet volume, $A_{\rm{la}} \propto -V$, giving: 
\begin{figure}
    \centering
    \includegraphics[width=\linewidth]{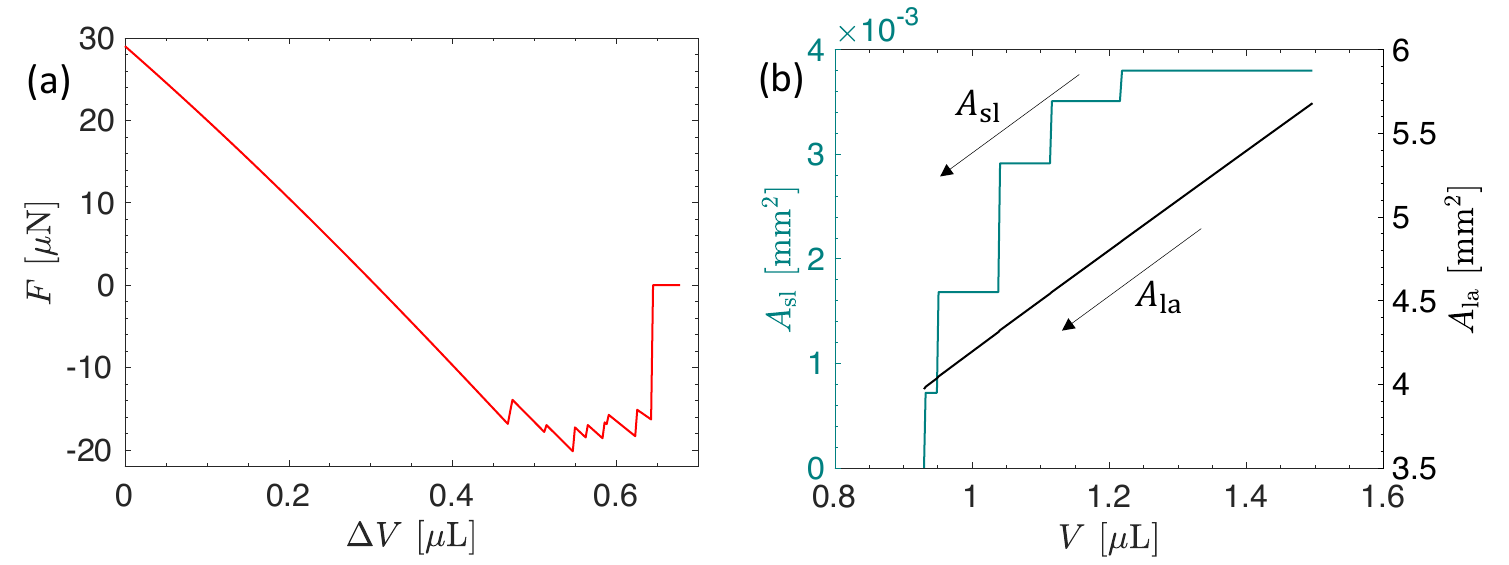} 
    \caption{(a) Simulation results of the variation in the net vertical force with the change in droplet volume during evaporation mode on a superhydrophobic surface with $\theta_{\rm{A}}=111.2$\textdegree, $\theta_{\rm{R}}=98.7$\tc and $\phi=0.05$. The droplet has an initial volume of $V=1.6$ $\mu$L. 
    (b) Variation in the solid-liquid ($A_{\rm{sl}}$) and liquid-air ($A_{\rm{la}}$) interfacial areas during the \textit{recede} stage in evaporation mode on a superhydrophobic surface with $\phi=0.016$, $\theta_{\rm{A}}=111.0$\tc and $\theta_{\rm{R}}=85.1$\textdegree. The droplet had an initial volume of 1.5 $\mu$L at the start of the \textit{recede} stage. This simulation replicates the experiments of \citeauthor{liimatainen2017mapping} \cite{liimatainen2017mapping}. }
    \label{fig:ras_area}
\end{figure}
%
%

\begin{equation}
\frac {dV}{dt} = \beta' V.   
 \label{eqn:time_scaling2}
\end{equation}
Here $\beta'$ is the new constant of proportionality that includes the area-volume scaling.

The relationship between the change in volume and the change in time can then be written as:
\begin{equation}
 \Delta t = \frac{\Delta V}{\beta'V} = \frac{\alpha}{\beta'} = C,  
 \label{eqn:time_scaling3}
\end{equation}
where $C$ is a constant.

 To plot the simulation data with the experimentally measured force values, we choose a suitable value of the constant $C$ by scaling the simulation time so that the initial slope of the curve $F$ versus $t$ during the start of the \textit{recede} stage is the same for both the experimental and the simulation data. This approach does not change to force magnitude or the relative position of different peaks and troughs in the sawtooth variation of the force.

\section{Experimental procedure}
\label{SI:experiments}
Direct force measurement experiments were performed using a DCAT-21 tensiometer (DataPhysics). A superhydrophobic pillared surface, with dimensions of $1.5 \times 1.5$ cm$^2$, was carefully positioned on a movable stage. A needle with an outer diameter of 0.91 mm was secured to a clamp connected to the tensiometer for force sensing. In this study, a DI water droplet (Milli-Q) of precisely 1.6 $\mu$L volume was dispensed using a syringe pump and then attached to the needle tip. To prevent water entrainment, the needle tip was sealed with Blu Tack. 

During the \textit{advance} stage, the stage was moved at a speed of 50 $\mu$m/s until the droplet contacted the surface. The upward motion of the stage was continued until the force increased to the reversal force ($F_{\rm{rev}} \sim 32$ $\mu$N). After reaching the reversal force, the stage was lowered at a speed of 25 $\mu$m/s until detachment of the droplet from the surface occurred. The interaction forces between the droplet and the surface were measured by the tensiometer at a frequency of 50 Hz. Simultaneously, the droplet dynamics were recorded using a DinoLite Edge Plus AM8917MZTL camera.

A second set of experiments were also conducted to study the droplet evaporation mode. In these experiments, the upward stage velocity was set to 10 $\mu$m/s while the downward velocity was set to zero.
\bibliography{ref}            

\begin{thebibliography}{97}%
\makeatletter
\providecommand \@ifxundefined [1]{%
 \@ifx{#1\undefined}
}%
\providecommand \@ifnum [1]{%
 \ifnum #1\expandafter \@firstoftwo
 \else \expandafter \@secondoftwo
 \fi
}%
\providecommand \@ifx [1]{%
 \ifx #1\expandafter \@firstoftwo
 \else \expandafter \@secondoftwo
 \fi
}%
\providecommand \natexlab [1]{#1}%
\providecommand \enquote  [1]{``#1''}%
\providecommand \bibnamefont  [1]{#1}%
\providecommand \bibfnamefont [1]{#1}%
\providecommand \citenamefont [1]{#1}%
\providecommand \href@noop [0]{\@secondoftwo}%
\providecommand \href [0]{\begingroup \@sanitize@url \@href}%
\providecommand \@href[1]{\@@startlink{#1}\@@href}%
\providecommand \@@href[1]{\endgroup#1\@@endlink}%
\providecommand \@sanitize@url [0]{\catcode `\\12\catcode `\$12\catcode
  `\&12\catcode `\#12\catcode `\^12\catcode `\_12\catcode `\%12\relax}%
\providecommand \@@startlink[1]{}%
\providecommand \@@endlink[0]{}%
\providecommand \url  [0]{\begingroup\@sanitize@url \@url }%
\providecommand \@url [1]{\endgroup\@href {#1}{\urlprefix }}%
\providecommand \urlprefix  [0]{URL }%
\providecommand \Eprint [0]{\href }%
\providecommand \doibase [0]{https://doi.org/}%
\providecommand \selectlanguage [0]{\@gobble}%
\providecommand \bibinfo  [0]{\@secondoftwo}%
\providecommand \bibfield  [0]{\@secondoftwo}%
\providecommand \translation [1]{[#1]}%
\providecommand \BibitemOpen [0]{}%
\providecommand \bibitemStop [0]{}%
\providecommand \bibitemNoStop [0]{.\EOS\space}%
\providecommand \EOS [0]{\spacefactor3000\relax}%
\providecommand \BibitemShut  [1]{\csname bibitem#1\endcsname}%
\let\auto@bib@innerbib\@empty
\bibitem [{\citenamefont {Barthlott}\ and\ \citenamefont
  {Neinhuis}(1997)}]{barthlott1997purity}%
  \BibitemOpen
  \bibfield  {author} {\bibinfo {author} {\bibfnamefont {W.}~\bibnamefont
  {Barthlott}}\ and\ \bibinfo {author} {\bibfnamefont {C.}~\bibnamefont
  {Neinhuis}},\ }\bibfield  {title} {\bibinfo {title} {Purity of the sacred
  lotus, or escape from contamination in biological surfaces},\ }\href@noop {}
  {\bibfield  {journal} {\bibinfo  {journal} {Planta}\ }\textbf {\bibinfo
  {volume} {202}},\ \bibinfo {pages} {1} (\bibinfo {year} {1997})}\BibitemShut
  {NoStop}%
\bibitem [{\citenamefont {Bhushan}\ and\ \citenamefont
  {Jung}(2011)}]{bhushan2011natural}%
  \BibitemOpen
  \bibfield  {author} {\bibinfo {author} {\bibfnamefont {B.}~\bibnamefont
  {Bhushan}}\ and\ \bibinfo {author} {\bibfnamefont {Y.~C.}\ \bibnamefont
  {Jung}},\ }\bibfield  {title} {\bibinfo {title} {Natural and biomimetic
  artificial surfaces for superhydrophobicity, self-cleaning, low adhesion, and
  drag reduction},\ }\href@noop {} {\bibfield  {journal} {\bibinfo  {journal}
  {Progress in Materials Science}\ }\textbf {\bibinfo {volume} {56}},\ \bibinfo
  {pages} {1} (\bibinfo {year} {2011})}\BibitemShut {NoStop}%
\bibitem [{\citenamefont {Nishimoto}\ and\ \citenamefont
  {Bhushan}(2013)}]{nishimoto2013bioinspired}%
  \BibitemOpen
  \bibfield  {author} {\bibinfo {author} {\bibfnamefont {S.}~\bibnamefont
  {Nishimoto}}\ and\ \bibinfo {author} {\bibfnamefont {B.}~\bibnamefont
  {Bhushan}},\ }\bibfield  {title} {\bibinfo {title} {Bioinspired self-cleaning
  surfaces with superhydrophobicity, superoleophobicity, and
  superhydrophilicity},\ }\href@noop {} {\bibfield  {journal} {\bibinfo
  {journal} {RSC Advances}\ }\textbf {\bibinfo {volume} {3}},\ \bibinfo {pages}
  {671} (\bibinfo {year} {2013})}\BibitemShut {NoStop}%
\bibitem [{\citenamefont {Feng}\ \emph {et~al.}(2008)\citenamefont {Feng},
  \citenamefont {Zhang}, \citenamefont {Xi}, \citenamefont {Zhu}, \citenamefont
  {Wang}, \citenamefont {Xia},\ and\ \citenamefont {Jiang}}]{feng2008petal}%
  \BibitemOpen
  \bibfield  {author} {\bibinfo {author} {\bibfnamefont {L.}~\bibnamefont
  {Feng}}, \bibinfo {author} {\bibfnamefont {Y.}~\bibnamefont {Zhang}},
  \bibinfo {author} {\bibfnamefont {J.}~\bibnamefont {Xi}}, \bibinfo {author}
  {\bibfnamefont {Y.}~\bibnamefont {Zhu}}, \bibinfo {author} {\bibfnamefont
  {N.}~\bibnamefont {Wang}}, \bibinfo {author} {\bibfnamefont {F.}~\bibnamefont
  {Xia}},\ and\ \bibinfo {author} {\bibfnamefont {L.}~\bibnamefont {Jiang}},\
  }\bibfield  {title} {\bibinfo {title} {Petal effect: a superhydrophobic state
  with high adhesive force},\ }\href@noop {} {\bibfield  {journal} {\bibinfo
  {journal} {Langmuir}\ }\textbf {\bibinfo {volume} {24}},\ \bibinfo {pages}
  {4114} (\bibinfo {year} {2008})}\BibitemShut {NoStop}%
\bibitem [{\citenamefont {Jiang}\ \emph {et~al.}(2024)\citenamefont {Jiang},
  \citenamefont {Xiao},\ and\ \citenamefont {Wei}}]{jiang2024sticky}%
  \BibitemOpen
  \bibfield  {author} {\bibinfo {author} {\bibfnamefont {Y.}~\bibnamefont
  {Jiang}}, \bibinfo {author} {\bibfnamefont {Y.}~\bibnamefont {Xiao}},\ and\
  \bibinfo {author} {\bibfnamefont {C.}~\bibnamefont {Wei}},\ }\bibfield
  {title} {\bibinfo {title} {Sticky superhydrophobic state},\ }\href@noop {}
  {\bibfield  {journal} {\bibinfo  {journal} {The Journal of Physical Chemistry
  Letters}\ }\textbf {\bibinfo {volume} {15}},\ \bibinfo {pages} {11896}
  (\bibinfo {year} {2024})}\BibitemShut {NoStop}%
\bibitem [{\citenamefont {Manna}\ and\ \citenamefont
  {Lynn}(2015)}]{manna2015fabrication}%
  \BibitemOpen
  \bibfield  {author} {\bibinfo {author} {\bibfnamefont {U.}~\bibnamefont
  {Manna}}\ and\ \bibinfo {author} {\bibfnamefont {D.~M.}\ \bibnamefont
  {Lynn}},\ }\bibfield  {title} {\bibinfo {title} {Fabrication of
  liquid-infused surfaces using reactive polymer multilayers: principles for
  manipulating the behaviors and mobilities of aqueous fluids on slippery
  liquid interfaces},\ }\href@noop {} {\bibfield  {journal} {\bibinfo
  {journal} {Advanced Materials}\ }\textbf {\bibinfo {volume} {27}},\ \bibinfo
  {pages} {3007} (\bibinfo {year} {2015})}\BibitemShut {NoStop}%
\bibitem [{\citenamefont {Peppou-Chapman}\ \emph {et~al.}(2020)\citenamefont
  {Peppou-Chapman}, \citenamefont {Hong}, \citenamefont {Waterhouse},\ and\
  \citenamefont {Neto}}]{peppou2020life}%
  \BibitemOpen
  \bibfield  {author} {\bibinfo {author} {\bibfnamefont {S.}~\bibnamefont
  {Peppou-Chapman}}, \bibinfo {author} {\bibfnamefont {J.~K.}\ \bibnamefont
  {Hong}}, \bibinfo {author} {\bibfnamefont {A.}~\bibnamefont {Waterhouse}},\
  and\ \bibinfo {author} {\bibfnamefont {C.}~\bibnamefont {Neto}},\ }\bibfield
  {title} {\bibinfo {title} {Life and death of liquid-infused surfaces: a
  review on the choice, analysis and fate of the infused liquid layer},\
  }\href@noop {} {\bibfield  {journal} {\bibinfo  {journal} {Chemical Society
  Reviews}\ }\textbf {\bibinfo {volume} {49}},\ \bibinfo {pages} {3688}
  (\bibinfo {year} {2020})}\BibitemShut {NoStop}%
\bibitem [{\citenamefont {Farhadi}\ \emph {et~al.}(2011)\citenamefont
  {Farhadi}, \citenamefont {Farzaneh},\ and\ \citenamefont
  {Kulinich}}]{farhadi2011anti}%
  \BibitemOpen
  \bibfield  {author} {\bibinfo {author} {\bibfnamefont {S.}~\bibnamefont
  {Farhadi}}, \bibinfo {author} {\bibfnamefont {M.}~\bibnamefont {Farzaneh}},\
  and\ \bibinfo {author} {\bibfnamefont {S.~A.}\ \bibnamefont {Kulinich}},\
  }\bibfield  {title} {\bibinfo {title} {Anti-icing performance of
  superhydrophobic surfaces},\ }\href@noop {} {\bibfield  {journal} {\bibinfo
  {journal} {Applied Surface Science}\ }\textbf {\bibinfo {volume} {257}},\
  \bibinfo {pages} {6264} (\bibinfo {year} {2011})}\BibitemShut {NoStop}%
\bibitem [{\citenamefont {Kreder}\ \emph {et~al.}(2016)\citenamefont {Kreder},
  \citenamefont {Alvarenga}, \citenamefont {Kim},\ and\ \citenamefont
  {Aizenberg}}]{kreder2016design}%
  \BibitemOpen
  \bibfield  {author} {\bibinfo {author} {\bibfnamefont {M.~J.}\ \bibnamefont
  {Kreder}}, \bibinfo {author} {\bibfnamefont {J.}~\bibnamefont {Alvarenga}},
  \bibinfo {author} {\bibfnamefont {P.}~\bibnamefont {Kim}},\ and\ \bibinfo
  {author} {\bibfnamefont {J.}~\bibnamefont {Aizenberg}},\ }\bibfield  {title}
  {\bibinfo {title} {Design of anti-icing surfaces: smooth, textured or
  slippery?},\ }\href@noop {} {\bibfield  {journal} {\bibinfo  {journal}
  {Nature Reviews Materials}\ }\textbf {\bibinfo {volume} {1}},\ \bibinfo
  {pages} {1} (\bibinfo {year} {2016})}\BibitemShut {NoStop}%
\bibitem [{\citenamefont {Amini}\ \emph {et~al.}(2017)\citenamefont {Amini},
  \citenamefont {Kolle}, \citenamefont {Petrone}, \citenamefont {Ahanotu},
  \citenamefont {Sunny}, \citenamefont {Sutanto}, \citenamefont {Hoon},
  \citenamefont {Cohen}, \citenamefont {Weaver}, \citenamefont {Aizenberg}
  \emph {et~al.}}]{amini2017preventing}%
  \BibitemOpen
  \bibfield  {author} {\bibinfo {author} {\bibfnamefont {S.}~\bibnamefont
  {Amini}}, \bibinfo {author} {\bibfnamefont {S.}~\bibnamefont {Kolle}},
  \bibinfo {author} {\bibfnamefont {L.}~\bibnamefont {Petrone}}, \bibinfo
  {author} {\bibfnamefont {O.}~\bibnamefont {Ahanotu}}, \bibinfo {author}
  {\bibfnamefont {S.}~\bibnamefont {Sunny}}, \bibinfo {author} {\bibfnamefont
  {C.~N.}\ \bibnamefont {Sutanto}}, \bibinfo {author} {\bibfnamefont
  {S.}~\bibnamefont {Hoon}}, \bibinfo {author} {\bibfnamefont {L.}~\bibnamefont
  {Cohen}}, \bibinfo {author} {\bibfnamefont {J.~C.}\ \bibnamefont {Weaver}},
  \bibinfo {author} {\bibfnamefont {J.}~\bibnamefont {Aizenberg}}, \emph
  {et~al.},\ }\bibfield  {title} {\bibinfo {title} {Preventing mussel adhesion
  using lubricant-infused materials},\ }\href@noop {} {\bibfield  {journal}
  {\bibinfo  {journal} {Science}\ }\textbf {\bibinfo {volume} {357}},\ \bibinfo
  {pages} {668} (\bibinfo {year} {2017})}\BibitemShut {NoStop}%
\bibitem [{\citenamefont {Sunny}\ \emph {et~al.}(2016)\citenamefont {Sunny},
  \citenamefont {Cheng}, \citenamefont {Daniel}, \citenamefont {Lo},
  \citenamefont {Ochoa}, \citenamefont {Howell}, \citenamefont {Vogel},
  \citenamefont {Majid},\ and\ \citenamefont
  {Aizenberg}}]{sunny2016transparent}%
  \BibitemOpen
  \bibfield  {author} {\bibinfo {author} {\bibfnamefont {S.}~\bibnamefont
  {Sunny}}, \bibinfo {author} {\bibfnamefont {G.}~\bibnamefont {Cheng}},
  \bibinfo {author} {\bibfnamefont {D.}~\bibnamefont {Daniel}}, \bibinfo
  {author} {\bibfnamefont {P.}~\bibnamefont {Lo}}, \bibinfo {author}
  {\bibfnamefont {S.}~\bibnamefont {Ochoa}}, \bibinfo {author} {\bibfnamefont
  {C.}~\bibnamefont {Howell}}, \bibinfo {author} {\bibfnamefont
  {N.}~\bibnamefont {Vogel}}, \bibinfo {author} {\bibfnamefont
  {A.}~\bibnamefont {Majid}},\ and\ \bibinfo {author} {\bibfnamefont
  {J.}~\bibnamefont {Aizenberg}},\ }\bibfield  {title} {\bibinfo {title}
  {Transparent antifouling material for improved operative field visibility in
  endoscopy},\ }\href@noop {} {\bibfield  {journal} {\bibinfo  {journal}
  {Proceedings of the National Academy of Sciences}\ }\textbf {\bibinfo
  {volume} {113}},\ \bibinfo {pages} {11676} (\bibinfo {year}
  {2016})}\BibitemShut {NoStop}%
\bibitem [{\citenamefont {Bocquet}\ and\ \citenamefont
  {Lauga}(2011)}]{bocquet2011smooth}%
  \BibitemOpen
  \bibfield  {author} {\bibinfo {author} {\bibfnamefont {L.}~\bibnamefont
  {Bocquet}}\ and\ \bibinfo {author} {\bibfnamefont {E.}~\bibnamefont
  {Lauga}},\ }\bibfield  {title} {\bibinfo {title} {A smooth future?},\
  }\href@noop {} {\bibfield  {journal} {\bibinfo  {journal} {Nature Materials}\
  }\textbf {\bibinfo {volume} {10}},\ \bibinfo {pages} {334} (\bibinfo {year}
  {2011})}\BibitemShut {NoStop}%
\bibitem [{\citenamefont {Xu}\ \emph {et~al.}(2020)\citenamefont {Xu},
  \citenamefont {Grabowski}, \citenamefont {Yu}, \citenamefont {Kerezyte},
  \citenamefont {Lee}, \citenamefont {Pfeifer},\ and\ \citenamefont
  {Kim}}]{xu2020superhydrophobic}%
  \BibitemOpen
  \bibfield  {author} {\bibinfo {author} {\bibfnamefont {M.}~\bibnamefont
  {Xu}}, \bibinfo {author} {\bibfnamefont {A.}~\bibnamefont {Grabowski}},
  \bibinfo {author} {\bibfnamefont {N.}~\bibnamefont {Yu}}, \bibinfo {author}
  {\bibfnamefont {G.}~\bibnamefont {Kerezyte}}, \bibinfo {author}
  {\bibfnamefont {J.-W.}\ \bibnamefont {Lee}}, \bibinfo {author} {\bibfnamefont
  {B.~R.}\ \bibnamefont {Pfeifer}},\ and\ \bibinfo {author} {\bibfnamefont
  {C.-J.}\ \bibnamefont {Kim}},\ }\bibfield  {title} {\bibinfo {title}
  {Superhydrophobic drag reduction for turbulent flows in open water},\
  }\href@noop {} {\bibfield  {journal} {\bibinfo  {journal} {Physical Review
  Applied}\ }\textbf {\bibinfo {volume} {13}},\ \bibinfo {pages} {034056}
  (\bibinfo {year} {2020})}\BibitemShut {NoStop}%
\bibitem [{\citenamefont {Zhang}\ \emph {et~al.}(2019)\citenamefont {Zhang},
  \citenamefont {Huang}, \citenamefont {Chen}, \citenamefont {Yang},\ and\
  \citenamefont {Lai}}]{zhang2019liquid}%
  \BibitemOpen
  \bibfield  {author} {\bibinfo {author} {\bibfnamefont {S.}~\bibnamefont
  {Zhang}}, \bibinfo {author} {\bibfnamefont {J.}~\bibnamefont {Huang}},
  \bibinfo {author} {\bibfnamefont {Z.}~\bibnamefont {Chen}}, \bibinfo {author}
  {\bibfnamefont {S.}~\bibnamefont {Yang}},\ and\ \bibinfo {author}
  {\bibfnamefont {Y.}~\bibnamefont {Lai}},\ }\bibfield  {title} {\bibinfo
  {title} {Liquid mobility on superwettable surfaces for applications in energy
  and the environment},\ }\href@noop {} {\bibfield  {journal} {\bibinfo
  {journal} {Journal of Materials Chemistry A}\ }\textbf {\bibinfo {volume}
  {7}},\ \bibinfo {pages} {38} (\bibinfo {year} {2019})}\BibitemShut {NoStop}%
\bibitem [{\citenamefont {Li}\ \emph {et~al.}(2019)\citenamefont {Li},
  \citenamefont {Fang}, \citenamefont {Li}, \citenamefont {Yang}, \citenamefont
  {Li}, \citenamefont {Li}, \citenamefont {Feng},\ and\ \citenamefont
  {Song}}]{li2019spontaneous}%
  \BibitemOpen
  \bibfield  {author} {\bibinfo {author} {\bibfnamefont {H.}~\bibnamefont
  {Li}}, \bibinfo {author} {\bibfnamefont {W.}~\bibnamefont {Fang}}, \bibinfo
  {author} {\bibfnamefont {Y.}~\bibnamefont {Li}}, \bibinfo {author}
  {\bibfnamefont {Q.}~\bibnamefont {Yang}}, \bibinfo {author} {\bibfnamefont
  {M.}~\bibnamefont {Li}}, \bibinfo {author} {\bibfnamefont {Q.}~\bibnamefont
  {Li}}, \bibinfo {author} {\bibfnamefont {X.-Q.}\ \bibnamefont {Feng}},\ and\
  \bibinfo {author} {\bibfnamefont {Y.}~\bibnamefont {Song}},\ }\bibfield
  {title} {\bibinfo {title} {Spontaneous droplets gyrating via asymmetric
  self-splitting on heterogeneous surfaces},\ }\href@noop {} {\bibfield
  {journal} {\bibinfo  {journal} {Nature Communications}\ }\textbf {\bibinfo
  {volume} {10}},\ \bibinfo {pages} {950} (\bibinfo {year} {2019})}\BibitemShut
  {NoStop}%
\bibitem [{\citenamefont {Bird}\ \emph {et~al.}(2013)\citenamefont {Bird},
  \citenamefont {Dhiman}, \citenamefont {Kwon},\ and\ \citenamefont
  {Varanasi}}]{bird2013reducing}%
  \BibitemOpen
  \bibfield  {author} {\bibinfo {author} {\bibfnamefont {J.~C.}\ \bibnamefont
  {Bird}}, \bibinfo {author} {\bibfnamefont {R.}~\bibnamefont {Dhiman}},
  \bibinfo {author} {\bibfnamefont {H.-M.}\ \bibnamefont {Kwon}},\ and\
  \bibinfo {author} {\bibfnamefont {K.~K.}\ \bibnamefont {Varanasi}},\
  }\bibfield  {title} {\bibinfo {title} {Reducing the contact time of a
  bouncing drop},\ }\href@noop {} {\bibfield  {journal} {\bibinfo  {journal}
  {Nature}\ }\textbf {\bibinfo {volume} {503}},\ \bibinfo {pages} {385}
  (\bibinfo {year} {2013})}\BibitemShut {NoStop}%
\bibitem [{\citenamefont {Villegas}\ \emph {et~al.}(2019)\citenamefont
  {Villegas}, \citenamefont {Zhang}, \citenamefont {Abu~Jarad}, \citenamefont
  {Soleymani},\ and\ \citenamefont {Didar}}]{villegas2019liquid}%
  \BibitemOpen
  \bibfield  {author} {\bibinfo {author} {\bibfnamefont {M.}~\bibnamefont
  {Villegas}}, \bibinfo {author} {\bibfnamefont {Y.}~\bibnamefont {Zhang}},
  \bibinfo {author} {\bibfnamefont {N.}~\bibnamefont {Abu~Jarad}}, \bibinfo
  {author} {\bibfnamefont {L.}~\bibnamefont {Soleymani}},\ and\ \bibinfo
  {author} {\bibfnamefont {T.~F.}\ \bibnamefont {Didar}},\ }\bibfield  {title}
  {\bibinfo {title} {Liquid-infused surfaces: a review of theory, design, and
  applications},\ }\href@noop {} {\bibfield  {journal} {\bibinfo  {journal}
  {Acs Nano}\ }\textbf {\bibinfo {volume} {13}},\ \bibinfo {pages} {8517}
  (\bibinfo {year} {2019})}\BibitemShut {NoStop}%
\bibitem [{\citenamefont {Qu{\'e}r{\'e}}(2005)}]{quere2005non}%
  \BibitemOpen
  \bibfield  {author} {\bibinfo {author} {\bibfnamefont {D.}~\bibnamefont
  {Qu{\'e}r{\'e}}},\ }\bibfield  {title} {\bibinfo {title} {Non-sticking
  drops},\ }\href@noop {} {\bibfield  {journal} {\bibinfo  {journal} {Reports
  on Progress in Physics}\ }\textbf {\bibinfo {volume} {68}},\ \bibinfo {pages}
  {2495} (\bibinfo {year} {2005})}\BibitemShut {NoStop}%
\bibitem [{\citenamefont {Roach}\ \emph {et~al.}(2008)\citenamefont {Roach},
  \citenamefont {Shirtcliffe},\ and\ \citenamefont
  {Newton}}]{roach2008progess}%
  \BibitemOpen
  \bibfield  {author} {\bibinfo {author} {\bibfnamefont {P.}~\bibnamefont
  {Roach}}, \bibinfo {author} {\bibfnamefont {N.~J.}\ \bibnamefont
  {Shirtcliffe}},\ and\ \bibinfo {author} {\bibfnamefont {M.~I.}\ \bibnamefont
  {Newton}},\ }\bibfield  {title} {\bibinfo {title} {Progess in
  superhydrophobic surface development},\ }\href@noop {} {\bibfield  {journal}
  {\bibinfo  {journal} {Soft Matter}\ }\textbf {\bibinfo {volume} {4}},\
  \bibinfo {pages} {224} (\bibinfo {year} {2008})}\BibitemShut {NoStop}%
\bibitem [{\citenamefont {Daniel}\ \emph {et~al.}(2023)\citenamefont {Daniel},
  \citenamefont {Vuckovac}, \citenamefont {Backholm}, \citenamefont {Latikka},
  \citenamefont {Karyappa}, \citenamefont {Koh}, \citenamefont {Timonen},
  \citenamefont {Tomczak},\ and\ \citenamefont {Ras}}]{daniel_probing_2023}%
  \BibitemOpen
  \bibfield  {author} {\bibinfo {author} {\bibfnamefont {D.}~\bibnamefont
  {Daniel}}, \bibinfo {author} {\bibfnamefont {M.}~\bibnamefont {Vuckovac}},
  \bibinfo {author} {\bibfnamefont {M.}~\bibnamefont {Backholm}}, \bibinfo
  {author} {\bibfnamefont {M.}~\bibnamefont {Latikka}}, \bibinfo {author}
  {\bibfnamefont {R.}~\bibnamefont {Karyappa}}, \bibinfo {author}
  {\bibfnamefont {X.~Q.}\ \bibnamefont {Koh}}, \bibinfo {author} {\bibfnamefont
  {J.~V.~I.}\ \bibnamefont {Timonen}}, \bibinfo {author} {\bibfnamefont
  {N.}~\bibnamefont {Tomczak}},\ and\ \bibinfo {author} {\bibfnamefont
  {R.~H.~A.}\ \bibnamefont {Ras}},\ }\bibfield  {title} {\bibinfo {title}
  {Probing surface wetting across multiple force, length and time scales},\
  }\href {https://doi.org/10.1038/s42005-023-01268-z} {\bibfield  {journal}
  {\bibinfo  {journal} {Communications Physics}\ }\textbf {\bibinfo {volume}
  {6}},\ \bibinfo {pages} {152} (\bibinfo {year} {2023})}\BibitemShut {NoStop}%
\bibitem [{\citenamefont {Wang}\ and\ \citenamefont
  {McCarthy}(2016)}]{wang2016covalently}%
  \BibitemOpen
  \bibfield  {author} {\bibinfo {author} {\bibfnamefont {L.}~\bibnamefont
  {Wang}}\ and\ \bibinfo {author} {\bibfnamefont {T.~J.}\ \bibnamefont
  {McCarthy}},\ }\bibfield  {title} {\bibinfo {title} {Covalently attached
  liquids: instant omniphobic surfaces with unprecedented repellency},\
  }\href@noop {} {\bibfield  {journal} {\bibinfo  {journal} {Angewandte Chemie
  International Edition}\ }\textbf {\bibinfo {volume} {55}},\ \bibinfo {pages}
  {244} (\bibinfo {year} {2016})}\BibitemShut {NoStop}%
\bibitem [{\citenamefont {Furmidge}(1962)}]{furmidge1962studies}%
  \BibitemOpen
  \bibfield  {author} {\bibinfo {author} {\bibfnamefont {C.}~\bibnamefont
  {Furmidge}},\ }\bibfield  {title} {\bibinfo {title} {Studies at phase
  interfaces. i. the sliding of liquid drops on solid surfaces and a theory for
  spray retention},\ }\href@noop {} {\bibfield  {journal} {\bibinfo  {journal}
  {Journal of Colloid Science}\ }\textbf {\bibinfo {volume} {17}},\ \bibinfo
  {pages} {309} (\bibinfo {year} {1962})}\BibitemShut {NoStop}%
\bibitem [{\citenamefont {Stern}\ \emph {et~al.}(2025)\citenamefont {Stern},
  \citenamefont {Tadmor}, \citenamefont {Miron},\ and\ \citenamefont
  {Vinod}}]{stern2025furmidge}%
  \BibitemOpen
  \bibfield  {author} {\bibinfo {author} {\bibfnamefont {Y.}~\bibnamefont
  {Stern}}, \bibinfo {author} {\bibfnamefont {R.}~\bibnamefont {Tadmor}},
  \bibinfo {author} {\bibfnamefont {A.}~\bibnamefont {Miron}},\ and\ \bibinfo
  {author} {\bibfnamefont {A.}~\bibnamefont {Vinod}},\ }\bibfield  {title}
  {\bibinfo {title} {Furmidge equation revisited},\ }\href@noop {} {\bibfield
  {journal} {\bibinfo  {journal} {Langmuir}\ }\textbf {\bibinfo {volume}
  {41}},\ \bibinfo {pages} {11785} (\bibinfo {year} {2025})}\BibitemShut
  {NoStop}%
\bibitem [{\citenamefont {McHale}\ \emph {et~al.}(2022)\citenamefont {McHale},
  \citenamefont {Gao}, \citenamefont {Wells}, \citenamefont {Barrio-Zhang},\
  and\ \citenamefont {Ledesma-Aguilar}}]{mchale2022friction}%
  \BibitemOpen
  \bibfield  {author} {\bibinfo {author} {\bibfnamefont {G.}~\bibnamefont
  {McHale}}, \bibinfo {author} {\bibfnamefont {N.}~\bibnamefont {Gao}},
  \bibinfo {author} {\bibfnamefont {G.~G.}\ \bibnamefont {Wells}}, \bibinfo
  {author} {\bibfnamefont {H.}~\bibnamefont {Barrio-Zhang}},\ and\ \bibinfo
  {author} {\bibfnamefont {R.}~\bibnamefont {Ledesma-Aguilar}},\ }\bibfield
  {title} {\bibinfo {title} {Friction coefficients for droplets on solids: The
  liquid--solid amontons’ laws},\ }\href@noop {} {\bibfield  {journal}
  {\bibinfo  {journal} {Langmuir}\ }\textbf {\bibinfo {volume} {38}},\ \bibinfo
  {pages} {4425} (\bibinfo {year} {2022})}\BibitemShut {NoStop}%
\bibitem [{\citenamefont {Gao}\ \emph {et~al.}(2018)\citenamefont {Gao},
  \citenamefont {Geyer}, \citenamefont {Pilat}, \citenamefont {Wooh},
  \citenamefont {Vollmer}, \citenamefont {Butt},\ and\ \citenamefont
  {Berger}}]{gao2018drops}%
  \BibitemOpen
  \bibfield  {author} {\bibinfo {author} {\bibfnamefont {N.}~\bibnamefont
  {Gao}}, \bibinfo {author} {\bibfnamefont {F.}~\bibnamefont {Geyer}}, \bibinfo
  {author} {\bibfnamefont {D.~W.}\ \bibnamefont {Pilat}}, \bibinfo {author}
  {\bibfnamefont {S.}~\bibnamefont {Wooh}}, \bibinfo {author} {\bibfnamefont
  {D.}~\bibnamefont {Vollmer}}, \bibinfo {author} {\bibfnamefont {H.-J.}\
  \bibnamefont {Butt}},\ and\ \bibinfo {author} {\bibfnamefont
  {R.}~\bibnamefont {Berger}},\ }\bibfield  {title} {\bibinfo {title} {How
  drops start sliding over solid surfaces},\ }\href@noop {} {\bibfield
  {journal} {\bibinfo  {journal} {Nature Physics}\ }\textbf {\bibinfo {volume}
  {14}},\ \bibinfo {pages} {191} (\bibinfo {year} {2018})}\BibitemShut
  {NoStop}%
\bibitem [{\citenamefont {Koh}\ \emph {et~al.}(2024)\citenamefont {Koh},
  \citenamefont {Thenarianto}, \citenamefont {Jokinen},\ and\ \citenamefont
  {Daniel}}]{koh2024quantifying}%
  \BibitemOpen
  \bibfield  {author} {\bibinfo {author} {\bibfnamefont {X.~Q.}\ \bibnamefont
  {Koh}}, \bibinfo {author} {\bibfnamefont {C.}~\bibnamefont {Thenarianto}},
  \bibinfo {author} {\bibfnamefont {V.}~\bibnamefont {Jokinen}},\ and\ \bibinfo
  {author} {\bibfnamefont {D.}~\bibnamefont {Daniel}},\ }\bibfield  {title}
  {\bibinfo {title} {Quantifying droplet--solid friction using an atomic force
  microscope},\ }\href@noop {} {\bibfield  {journal} {\bibinfo  {journal}
  {Droplet}\ }\textbf {\bibinfo {volume} {3}},\ \bibinfo {pages} {e107}
  (\bibinfo {year} {2024})}\BibitemShut {NoStop}%
\bibitem [{\citenamefont {Allred}\ \emph {et~al.}(2017)\citenamefont {Allred},
  \citenamefont {Weibel},\ and\ \citenamefont
  {Garimella}}]{allred2017wettability}%
  \BibitemOpen
  \bibfield  {author} {\bibinfo {author} {\bibfnamefont {T.~P.}\ \bibnamefont
  {Allred}}, \bibinfo {author} {\bibfnamefont {J.~A.}\ \bibnamefont {Weibel}},\
  and\ \bibinfo {author} {\bibfnamefont {S.~V.}\ \bibnamefont {Garimella}},\
  }\bibfield  {title} {\bibinfo {title} {A wettability metric for
  characterization of capillary flow on textured superhydrophilic surfaces},\
  }\href@noop {} {\bibfield  {journal} {\bibinfo  {journal} {Langmuir}\
  }\textbf {\bibinfo {volume} {33}},\ \bibinfo {pages} {7847} (\bibinfo {year}
  {2017})}\BibitemShut {NoStop}%
\bibitem [{\citenamefont {Huhtam{\"a}ki}\ \emph {et~al.}(2018)\citenamefont
  {Huhtam{\"a}ki}, \citenamefont {Tian}, \citenamefont {Korhonen},\ and\
  \citenamefont {Ras}}]{huhtamaki2018surface}%
  \BibitemOpen
  \bibfield  {author} {\bibinfo {author} {\bibfnamefont {T.}~\bibnamefont
  {Huhtam{\"a}ki}}, \bibinfo {author} {\bibfnamefont {X.}~\bibnamefont {Tian}},
  \bibinfo {author} {\bibfnamefont {J.~T.}\ \bibnamefont {Korhonen}},\ and\
  \bibinfo {author} {\bibfnamefont {R.~H.}\ \bibnamefont {Ras}},\ }\bibfield
  {title} {\bibinfo {title} {Surface-wetting characterization using
  contact-angle measurements},\ }\href@noop {} {\bibfield  {journal} {\bibinfo
  {journal} {Nature Protocols}\ }\textbf {\bibinfo {volume} {13}},\ \bibinfo
  {pages} {1521} (\bibinfo {year} {2018})}\BibitemShut {NoStop}%
\bibitem [{\citenamefont {Daniel}\ \emph
  {et~al.}(2019{\natexlab{a}})\citenamefont {Daniel}, \citenamefont {Lay},
  \citenamefont {Sng}, \citenamefont {Jun~Lee}, \citenamefont {Jin~Neo},
  \citenamefont {Ling},\ and\ \citenamefont {Tomczak}}]{daniel_mapping_2019}%
  \BibitemOpen
  \bibfield  {author} {\bibinfo {author} {\bibfnamefont {D.}~\bibnamefont
  {Daniel}}, \bibinfo {author} {\bibfnamefont {C.~L.}\ \bibnamefont {Lay}},
  \bibinfo {author} {\bibfnamefont {A.}~\bibnamefont {Sng}}, \bibinfo {author}
  {\bibfnamefont {C.~J.}\ \bibnamefont {Jun~Lee}}, \bibinfo {author}
  {\bibfnamefont {D.~C.}\ \bibnamefont {Jin~Neo}}, \bibinfo {author}
  {\bibfnamefont {X.~Y.}\ \bibnamefont {Ling}},\ and\ \bibinfo {author}
  {\bibfnamefont {N.}~\bibnamefont {Tomczak}},\ }\bibfield  {title} {\bibinfo
  {title} {Mapping micrometer-scale wetting properties of superhydrophobic
  surfaces},\ }\href {https://doi.org/10.1073/pnas.1916772116} {\bibfield
  {journal} {\bibinfo  {journal} {Proceedings of the National Academy of
  Sciences}\ }\textbf {\bibinfo {volume} {116}},\ \bibinfo {pages} {25008}
  (\bibinfo {year} {2019}{\natexlab{a}})},\ \bibinfo {note} {publisher:
  Proceedings of the National Academy of Sciences}\BibitemShut {NoStop}%
\bibitem [{\citenamefont {Kumar}\ \emph {et~al.}(2025)\citenamefont {Kumar},
  \citenamefont {Krasowska},\ and\ \citenamefont {Berry}}]{kumar2025adhesion}%
  \BibitemOpen
  \bibfield  {author} {\bibinfo {author} {\bibfnamefont {P.}~\bibnamefont
  {Kumar}}, \bibinfo {author} {\bibfnamefont {M.}~\bibnamefont {Krasowska}},\
  and\ \bibinfo {author} {\bibfnamefont {J.~D.}\ \bibnamefont {Berry}},\ }\href
  {https://arxiv.org/abs/2511.19908} {\bibinfo {title} {Probing
  superhydrophobic surface topography using droplet adhesion}} (\bibinfo {year}
  {2025}),\ \bibinfo {note} {manuscript under review},\ \Eprint
  {https://arxiv.org/abs/2511.19908} {arXiv:2511.19908 [physics.flu-dyn]}
  \BibitemShut {NoStop}%
\bibitem [{\citenamefont {Daniel}\ \emph {et~al.}(2020)\citenamefont {Daniel},
  \citenamefont {Florida}, \citenamefont {Lay}, \citenamefont {Koh},
  \citenamefont {Sng},\ and\ \citenamefont
  {Tomczak}}]{daniel_quantifying_2020}%
  \BibitemOpen
  \bibfield  {author} {\bibinfo {author} {\bibfnamefont {D.}~\bibnamefont
  {Daniel}}, \bibinfo {author} {\bibfnamefont {Y.}~\bibnamefont {Florida}},
  \bibinfo {author} {\bibfnamefont {C.~L.}\ \bibnamefont {Lay}}, \bibinfo
  {author} {\bibfnamefont {X.~Q.}\ \bibnamefont {Koh}}, \bibinfo {author}
  {\bibfnamefont {A.}~\bibnamefont {Sng}},\ and\ \bibinfo {author}
  {\bibfnamefont {N.}~\bibnamefont {Tomczak}},\ }\bibfield  {title} {\bibinfo
  {title} {Quantifying {Surface} {Wetting} {Properties} {Using} {Droplet}
  {Probe} {Atomic} {Force} {Microscopy}},\ }\href
  {https://doi.org/10.1021/acsami.0c12123} {\bibfield  {journal} {\bibinfo
  {journal} {ACS Applied Materials \& Interfaces}\ }\textbf {\bibinfo {volume}
  {12}},\ \bibinfo {pages} {42386} (\bibinfo {year} {2020})}\BibitemShut
  {NoStop}%
\bibitem [{\citenamefont {Liimatainen}\ \emph {et~al.}(2017)\citenamefont
  {Liimatainen}, \citenamefont {Vuckovac}, \citenamefont {Jokinen},
  \citenamefont {Sariola}, \citenamefont {Hokkanen}, \citenamefont {Zhou},\
  and\ \citenamefont {Ras}}]{liimatainen2017mapping}%
  \BibitemOpen
  \bibfield  {author} {\bibinfo {author} {\bibfnamefont {V.}~\bibnamefont
  {Liimatainen}}, \bibinfo {author} {\bibfnamefont {M.}~\bibnamefont
  {Vuckovac}}, \bibinfo {author} {\bibfnamefont {V.}~\bibnamefont {Jokinen}},
  \bibinfo {author} {\bibfnamefont {V.}~\bibnamefont {Sariola}}, \bibinfo
  {author} {\bibfnamefont {M.~J.}\ \bibnamefont {Hokkanen}}, \bibinfo {author}
  {\bibfnamefont {Q.}~\bibnamefont {Zhou}},\ and\ \bibinfo {author}
  {\bibfnamefont {R.~H.}\ \bibnamefont {Ras}},\ }\bibfield  {title} {\bibinfo
  {title} {Mapping microscale wetting variations on biological and synthetic
  water-repellent surfaces},\ }\href@noop {} {\bibfield  {journal} {\bibinfo
  {journal} {Nature Communications}\ }\textbf {\bibinfo {volume} {8}},\
  \bibinfo {pages} {1798} (\bibinfo {year} {2017})}\BibitemShut {NoStop}%
\bibitem [{\citenamefont {Hokkanen}\ \emph {et~al.}(2021)\citenamefont
  {Hokkanen}, \citenamefont {Backholm}, \citenamefont {Vuckovac}, \citenamefont
  {Zhou},\ and\ \citenamefont {Ras}}]{hokkanen_forcebased_2021}%
  \BibitemOpen
  \bibfield  {author} {\bibinfo {author} {\bibfnamefont {M.~J.}\ \bibnamefont
  {Hokkanen}}, \bibinfo {author} {\bibfnamefont {M.}~\bibnamefont {Backholm}},
  \bibinfo {author} {\bibfnamefont {M.}~\bibnamefont {Vuckovac}}, \bibinfo
  {author} {\bibfnamefont {Q.}~\bibnamefont {Zhou}},\ and\ \bibinfo {author}
  {\bibfnamefont {R.~H.~A.}\ \bibnamefont {Ras}},\ }\bibfield  {title}
  {\bibinfo {title} {Force‐{Based} {Wetting} {Characterization} of
  {Stochastic} {Superhydrophobic} {Coatings} at {Nanonewton} {Sensitivity}},\
  }\href {https://doi.org/10.1002/adma.202105130} {\bibfield  {journal}
  {\bibinfo  {journal} {Advanced Materials}\ }\textbf {\bibinfo {volume}
  {33}},\ \bibinfo {pages} {2105130} (\bibinfo {year} {2021})}\BibitemShut
  {NoStop}%
\bibitem [{\citenamefont {Pilat}\ \emph {et~al.}(2012)\citenamefont {Pilat},
  \citenamefont {Papadopoulos}, \citenamefont {Schäffel}, \citenamefont
  {Vollmer}, \citenamefont {Berger},\ and\ \citenamefont
  {Butt}}]{pilat_dynamic_2012}%
  \BibitemOpen
  \bibfield  {author} {\bibinfo {author} {\bibfnamefont {D.~W.}\ \bibnamefont
  {Pilat}}, \bibinfo {author} {\bibfnamefont {P.}~\bibnamefont {Papadopoulos}},
  \bibinfo {author} {\bibfnamefont {D.}~\bibnamefont {Schäffel}}, \bibinfo
  {author} {\bibfnamefont {D.}~\bibnamefont {Vollmer}}, \bibinfo {author}
  {\bibfnamefont {R.}~\bibnamefont {Berger}},\ and\ \bibinfo {author}
  {\bibfnamefont {H.-J.}\ \bibnamefont {Butt}},\ }\bibfield  {title} {\bibinfo
  {title} {Dynamic {Measurement} of the {Force} {Required} to {Move} a {Liquid}
  {Drop} on a {Solid} {Surface}},\ }\href {https://doi.org/10.1021/la3041067}
  {\bibfield  {journal} {\bibinfo  {journal} {Langmuir}\ }\textbf {\bibinfo
  {volume} {28}},\ \bibinfo {pages} {16812} (\bibinfo {year}
  {2012})}\BibitemShut {NoStop}%
\bibitem [{\citenamefont {Ko}\ \emph {et~al.}(2020)\citenamefont {Ko},
  \citenamefont {Park}, \citenamefont {Kim}, \citenamefont {Yoon},
  \citenamefont {Kim}, \citenamefont {Oh}, \citenamefont {Nahm},\ and\
  \citenamefont {Moon}}]{ko2020single}%
  \BibitemOpen
  \bibfield  {author} {\bibinfo {author} {\bibfnamefont {T.-J.}\ \bibnamefont
  {Ko}}, \bibinfo {author} {\bibfnamefont {S.~J.}\ \bibnamefont {Park}},
  \bibinfo {author} {\bibfnamefont {M.-S.}\ \bibnamefont {Kim}}, \bibinfo
  {author} {\bibfnamefont {S.~M.}\ \bibnamefont {Yoon}}, \bibinfo {author}
  {\bibfnamefont {S.~J.}\ \bibnamefont {Kim}}, \bibinfo {author} {\bibfnamefont
  {K.~H.}\ \bibnamefont {Oh}}, \bibinfo {author} {\bibfnamefont
  {S.}~\bibnamefont {Nahm}},\ and\ \bibinfo {author} {\bibfnamefont {M.-W.}\
  \bibnamefont {Moon}},\ }\bibfield  {title} {\bibinfo {title} {Single-step
  plasma-induced hierarchical structures for tunable water adhesion},\
  }\href@noop {} {\bibfield  {journal} {\bibinfo  {journal} {Scientific
  reports}\ }\textbf {\bibinfo {volume} {10}},\ \bibinfo {pages} {874}
  (\bibinfo {year} {2020})}\BibitemShut {NoStop}%
\bibitem [{\citenamefont {Shi}\ \emph {et~al.}(2015)\citenamefont {Shi},
  \citenamefont {Cui}, \citenamefont {Xie}, \citenamefont {Liu}, \citenamefont
  {Chan}, \citenamefont {Israelachvili},\ and\ \citenamefont
  {Zeng}}]{shi_measuring_2015}%
  \BibitemOpen
  \bibfield  {author} {\bibinfo {author} {\bibfnamefont {C.}~\bibnamefont
  {Shi}}, \bibinfo {author} {\bibfnamefont {X.}~\bibnamefont {Cui}}, \bibinfo
  {author} {\bibfnamefont {L.}~\bibnamefont {Xie}}, \bibinfo {author}
  {\bibfnamefont {Q.}~\bibnamefont {Liu}}, \bibinfo {author} {\bibfnamefont
  {D.~Y.~C.}\ \bibnamefont {Chan}}, \bibinfo {author} {\bibfnamefont {J.~N.}\
  \bibnamefont {Israelachvili}},\ and\ \bibinfo {author} {\bibfnamefont
  {H.}~\bibnamefont {Zeng}},\ }\bibfield  {title} {\bibinfo {title} {Measuring
  {Forces} and {Spatiotemporal} {Evolution} of {Thin} {Water} {Films} between
  an {Air} {Bubble} and {Solid} {Surfaces} of {Different} {Hydrophobicity}},\
  }\href {https://doi.org/10.1021/nn506601j} {\bibfield  {journal} {\bibinfo
  {journal} {ACS Nano}\ }\textbf {\bibinfo {volume} {9}},\ \bibinfo {pages}
  {95} (\bibinfo {year} {2015})}\BibitemShut {NoStop}%
\bibitem [{\citenamefont {Daniel}\ \emph {et~al.}(2017)\citenamefont {Daniel},
  \citenamefont {Timonen}, \citenamefont {Li}, \citenamefont {Velling},\ and\
  \citenamefont {Aizenberg}}]{daniel2017oleoplaning}%
  \BibitemOpen
  \bibfield  {author} {\bibinfo {author} {\bibfnamefont {D.}~\bibnamefont
  {Daniel}}, \bibinfo {author} {\bibfnamefont {J.~V.}\ \bibnamefont {Timonen}},
  \bibinfo {author} {\bibfnamefont {R.}~\bibnamefont {Li}}, \bibinfo {author}
  {\bibfnamefont {S.~J.}\ \bibnamefont {Velling}},\ and\ \bibinfo {author}
  {\bibfnamefont {J.}~\bibnamefont {Aizenberg}},\ }\bibfield  {title} {\bibinfo
  {title} {Oleoplaning droplets on lubricated surfaces},\ }\href@noop {}
  {\bibfield  {journal} {\bibinfo  {journal} {Nature Physics}\ }\textbf
  {\bibinfo {volume} {13}},\ \bibinfo {pages} {1020} (\bibinfo {year}
  {2017})}\BibitemShut {NoStop}%
\bibitem [{\citenamefont {Daniel}\ \emph
  {et~al.}(2019{\natexlab{b}})\citenamefont {Daniel}, \citenamefont {Chia},
  \citenamefont {Moh}, \citenamefont {Liu}, \citenamefont {Koh}, \citenamefont
  {Zhang},\ and\ \citenamefont {Tomczak}}]{daniel2019hydration}%
  \BibitemOpen
  \bibfield  {author} {\bibinfo {author} {\bibfnamefont {D.}~\bibnamefont
  {Daniel}}, \bibinfo {author} {\bibfnamefont {A.~Y.~T.}\ \bibnamefont {Chia}},
  \bibinfo {author} {\bibfnamefont {L.~C.~H.}\ \bibnamefont {Moh}}, \bibinfo
  {author} {\bibfnamefont {R.}~\bibnamefont {Liu}}, \bibinfo {author}
  {\bibfnamefont {X.~Q.}\ \bibnamefont {Koh}}, \bibinfo {author} {\bibfnamefont
  {X.}~\bibnamefont {Zhang}},\ and\ \bibinfo {author} {\bibfnamefont
  {N.}~\bibnamefont {Tomczak}},\ }\bibfield  {title} {\bibinfo {title}
  {Hydration lubrication of polyzwitterionic brushes leads to nearly
  friction-and adhesion-free droplet motion},\ }\href@noop {} {\bibfield
  {journal} {\bibinfo  {journal} {Communications Physics}\ }\textbf {\bibinfo
  {volume} {2}},\ \bibinfo {pages} {105} (\bibinfo {year}
  {2019}{\natexlab{b}})}\BibitemShut {NoStop}%
\bibitem [{\citenamefont {Backholm}\ \emph {et~al.}(2020)\citenamefont
  {Backholm}, \citenamefont {Molpeceres}, \citenamefont {Vuckovac},
  \citenamefont {Nurmi}, \citenamefont {Hokkanen}, \citenamefont {Jokinen},
  \citenamefont {Timonen},\ and\ \citenamefont {Ras}}]{backholm2020water}%
  \BibitemOpen
  \bibfield  {author} {\bibinfo {author} {\bibfnamefont {M.}~\bibnamefont
  {Backholm}}, \bibinfo {author} {\bibfnamefont {D.}~\bibnamefont
  {Molpeceres}}, \bibinfo {author} {\bibfnamefont {M.}~\bibnamefont
  {Vuckovac}}, \bibinfo {author} {\bibfnamefont {H.}~\bibnamefont {Nurmi}},
  \bibinfo {author} {\bibfnamefont {M.~J.}\ \bibnamefont {Hokkanen}}, \bibinfo
  {author} {\bibfnamefont {V.}~\bibnamefont {Jokinen}}, \bibinfo {author}
  {\bibfnamefont {J.~V.}\ \bibnamefont {Timonen}},\ and\ \bibinfo {author}
  {\bibfnamefont {R.~H.}\ \bibnamefont {Ras}},\ }\bibfield  {title} {\bibinfo
  {title} {Water droplet friction and rolling dynamics on superhydrophobic
  surfaces},\ }\href@noop {} {\bibfield  {journal} {\bibinfo  {journal}
  {Communications Materials}\ }\textbf {\bibinfo {volume} {1}},\ \bibinfo
  {pages} {64} (\bibinfo {year} {2020})}\BibitemShut {NoStop}%
\bibitem [{\citenamefont {Hinduja}\ \emph {et~al.}(2022)\citenamefont
  {Hinduja}, \citenamefont {Laroche}, \citenamefont {Shumaly}, \citenamefont
  {Wang}, \citenamefont {Vollmer}, \citenamefont {Butt},\ and\ \citenamefont
  {Berger}}]{hinduja2022scanning}%
  \BibitemOpen
  \bibfield  {author} {\bibinfo {author} {\bibfnamefont {C.}~\bibnamefont
  {Hinduja}}, \bibinfo {author} {\bibfnamefont {A.}~\bibnamefont {Laroche}},
  \bibinfo {author} {\bibfnamefont {S.}~\bibnamefont {Shumaly}}, \bibinfo
  {author} {\bibfnamefont {Y.}~\bibnamefont {Wang}}, \bibinfo {author}
  {\bibfnamefont {D.}~\bibnamefont {Vollmer}}, \bibinfo {author} {\bibfnamefont
  {H.-J.}\ \bibnamefont {Butt}},\ and\ \bibinfo {author} {\bibfnamefont
  {R.}~\bibnamefont {Berger}},\ }\bibfield  {title} {\bibinfo {title} {Scanning
  drop friction force microscopy},\ }\href@noop {} {\bibfield  {journal}
  {\bibinfo  {journal} {Langmuir}\ }\textbf {\bibinfo {volume} {38}},\ \bibinfo
  {pages} {14635} (\bibinfo {year} {2022})}\BibitemShut {NoStop}%
\bibitem [{\citenamefont {Escobar}\ \emph {et~al.}(2017)\citenamefont
  {Escobar}, \citenamefont {Garza},\ and\ \citenamefont
  {Castillo}}]{escobar2017measuring}%
  \BibitemOpen
  \bibfield  {author} {\bibinfo {author} {\bibfnamefont {J.~V.}\ \bibnamefont
  {Escobar}}, \bibinfo {author} {\bibfnamefont {C.}~\bibnamefont {Garza}},\
  and\ \bibinfo {author} {\bibfnamefont {R.}~\bibnamefont {Castillo}},\
  }\bibfield  {title} {\bibinfo {title} {Measuring adhesion on rough surfaces
  using atomic force microscopy with a liquid probe},\ }\href@noop {}
  {\bibfield  {journal} {\bibinfo  {journal} {Beilstein Journal of
  Nanotechnology}\ }\textbf {\bibinfo {volume} {8}},\ \bibinfo {pages} {813}
  (\bibinfo {year} {2017})}\BibitemShut {NoStop}%
\bibitem [{\citenamefont {Shi}\ \emph {et~al.}(2016)\citenamefont {Shi},
  \citenamefont {Yan}, \citenamefont {Xie}, \citenamefont {Zhang},
  \citenamefont {Wang}, \citenamefont {Takahara},\ and\ \citenamefont
  {Zeng}}]{shi2016long}%
  \BibitemOpen
  \bibfield  {author} {\bibinfo {author} {\bibfnamefont {C.}~\bibnamefont
  {Shi}}, \bibinfo {author} {\bibfnamefont {B.}~\bibnamefont {Yan}}, \bibinfo
  {author} {\bibfnamefont {L.}~\bibnamefont {Xie}}, \bibinfo {author}
  {\bibfnamefont {L.}~\bibnamefont {Zhang}}, \bibinfo {author} {\bibfnamefont
  {J.}~\bibnamefont {Wang}}, \bibinfo {author} {\bibfnamefont {A.}~\bibnamefont
  {Takahara}},\ and\ \bibinfo {author} {\bibfnamefont {H.}~\bibnamefont
  {Zeng}},\ }\bibfield  {title} {\bibinfo {title} {Long-range hydrophilic
  attraction between water and polyelectrolyte surfaces in oil},\ }\href@noop
  {} {\bibfield  {journal} {\bibinfo  {journal} {Angewandte Chemie
  International Edition}\ }\textbf {\bibinfo {volume} {55}},\ \bibinfo {pages}
  {15017} (\bibinfo {year} {2016})}\BibitemShut {NoStop}%
\bibitem [{\citenamefont {Samuel}\ \emph {et~al.}(2011)\citenamefont {Samuel},
  \citenamefont {Zhao},\ and\ \citenamefont {Law}}]{samuel2011study}%
  \BibitemOpen
  \bibfield  {author} {\bibinfo {author} {\bibfnamefont {B.}~\bibnamefont
  {Samuel}}, \bibinfo {author} {\bibfnamefont {H.}~\bibnamefont {Zhao}},\ and\
  \bibinfo {author} {\bibfnamefont {K.-Y.}\ \bibnamefont {Law}},\ }\bibfield
  {title} {\bibinfo {title} {Study of wetting and adhesion interactions between
  water and various polymer and superhydrophobic surfaces},\ }\href@noop {}
  {\bibfield  {journal} {\bibinfo  {journal} {The Journal of Physical Chemistry
  C}\ }\textbf {\bibinfo {volume} {115}},\ \bibinfo {pages} {14852} (\bibinfo
  {year} {2011})}\BibitemShut {NoStop}%
\bibitem [{\citenamefont {Livi}\ \emph {et~al.}(2025)\citenamefont {Livi},
  \citenamefont {Fazal}, \citenamefont {Barrio-Zhang}, \citenamefont {Zhao},
  \citenamefont {McHale}, \citenamefont {Wells}, \citenamefont {Syed},
  \citenamefont {Radacsi},\ and\ \citenamefont
  {Koutsos}}]{livi2025characterization}%
  \BibitemOpen
  \bibfield  {author} {\bibinfo {author} {\bibfnamefont {G.}~\bibnamefont
  {Livi}}, \bibinfo {author} {\bibfnamefont {F.}~\bibnamefont {Fazal}},
  \bibinfo {author} {\bibfnamefont {H.}~\bibnamefont {Barrio-Zhang}}, \bibinfo
  {author} {\bibfnamefont {H.}~\bibnamefont {Zhao}}, \bibinfo {author}
  {\bibfnamefont {G.}~\bibnamefont {McHale}}, \bibinfo {author} {\bibfnamefont
  {G.~G.}\ \bibnamefont {Wells}}, \bibinfo {author} {\bibfnamefont
  {A.}~\bibnamefont {Syed}}, \bibinfo {author} {\bibfnamefont {N.}~\bibnamefont
  {Radacsi}},\ and\ \bibinfo {author} {\bibfnamefont {V.}~\bibnamefont
  {Koutsos}},\ }\bibfield  {title} {\bibinfo {title} {Characterization of water
  adhesion of pcl and plcl electrospun fiber mats and their correlation to
  wettability},\ }\href@noop {} {\bibfield  {journal} {\bibinfo  {journal}
  {Macromolecular Materials and Engineering}\ ,\ \bibinfo {pages} {e00189}}
  (\bibinfo {year} {2025})}\BibitemShut {NoStop}%
\bibitem [{\citenamefont {Lepikko}\ \emph {et~al.}(2025)\citenamefont
  {Lepikko}, \citenamefont {Turkki}, \citenamefont {Koskinen}, \citenamefont
  {Raju}, \citenamefont {Jokinen}, \citenamefont {Kiseleva}, \citenamefont
  {Rantataro}, \citenamefont {Timonen}, \citenamefont {Backholm}, \citenamefont
  {Tittonen} \emph {et~al.}}]{lepikko2025droplet}%
  \BibitemOpen
  \bibfield  {author} {\bibinfo {author} {\bibfnamefont {S.}~\bibnamefont
  {Lepikko}}, \bibinfo {author} {\bibfnamefont {V.}~\bibnamefont {Turkki}},
  \bibinfo {author} {\bibfnamefont {T.}~\bibnamefont {Koskinen}}, \bibinfo
  {author} {\bibfnamefont {R.}~\bibnamefont {Raju}}, \bibinfo {author}
  {\bibfnamefont {V.}~\bibnamefont {Jokinen}}, \bibinfo {author} {\bibfnamefont
  {M.~S.}\ \bibnamefont {Kiseleva}}, \bibinfo {author} {\bibfnamefont
  {S.}~\bibnamefont {Rantataro}}, \bibinfo {author} {\bibfnamefont {J.~V.}\
  \bibnamefont {Timonen}}, \bibinfo {author} {\bibfnamefont {M.}~\bibnamefont
  {Backholm}}, \bibinfo {author} {\bibfnamefont {I.}~\bibnamefont {Tittonen}},
  \emph {et~al.},\ }\bibfield  {title} {\bibinfo {title} {Droplet friction on
  superhydrophobic surfaces scales with liquid-solid contact fraction},\
  }\href@noop {} {\bibfield  {journal} {\bibinfo  {journal} {Small}\ }\textbf
  {\bibinfo {volume} {21}},\ \bibinfo {pages} {2405335} (\bibinfo {year}
  {2025})}\BibitemShut {NoStop}%
\bibitem [{\citenamefont {Backholm}\ \emph {et~al.}(2024)\citenamefont
  {Backholm}, \citenamefont {K{\"a}rki}, \citenamefont {Nurmi}, \citenamefont
  {Vuckovac}, \citenamefont {Turkki}, \citenamefont {Lepikko}, \citenamefont
  {Jokinen}, \citenamefont {Qu{\'e}r{\'e}}, \citenamefont {Timonen},\ and\
  \citenamefont {Ras}}]{backholm2024toward}%
  \BibitemOpen
  \bibfield  {author} {\bibinfo {author} {\bibfnamefont {M.}~\bibnamefont
  {Backholm}}, \bibinfo {author} {\bibfnamefont {T.}~\bibnamefont {K{\"a}rki}},
  \bibinfo {author} {\bibfnamefont {H.~A.}\ \bibnamefont {Nurmi}}, \bibinfo
  {author} {\bibfnamefont {M.}~\bibnamefont {Vuckovac}}, \bibinfo {author}
  {\bibfnamefont {V.}~\bibnamefont {Turkki}}, \bibinfo {author} {\bibfnamefont
  {S.}~\bibnamefont {Lepikko}}, \bibinfo {author} {\bibfnamefont
  {V.}~\bibnamefont {Jokinen}}, \bibinfo {author} {\bibfnamefont
  {D.}~\bibnamefont {Qu{\'e}r{\'e}}}, \bibinfo {author} {\bibfnamefont {J.~V.}\
  \bibnamefont {Timonen}},\ and\ \bibinfo {author} {\bibfnamefont {R.~H.}\
  \bibnamefont {Ras}},\ }\bibfield  {title} {\bibinfo {title} {Toward vanishing
  droplet friction on repellent surfaces},\ }\href@noop {} {\bibfield
  {journal} {\bibinfo  {journal} {Proceedings of the National Academy of
  Sciences}\ }\textbf {\bibinfo {volume} {121}},\ \bibinfo {pages}
  {e2315214121} (\bibinfo {year} {2024})}\BibitemShut {NoStop}%
\bibitem [{\citenamefont {Nagy}(2024)}]{nagy2024determination}%
  \BibitemOpen
  \bibfield  {author} {\bibinfo {author} {\bibfnamefont {N.}~\bibnamefont
  {Nagy}},\ }\bibfield  {title} {\bibinfo {title} {Determination of
  solid-liquid adhesion work on flat surfaces in a direct and absolute
  manner},\ }\href@noop {} {\bibfield  {journal} {\bibinfo  {journal}
  {Scientific Reports}\ }\textbf {\bibinfo {volume} {14}},\ \bibinfo {pages}
  {29991} (\bibinfo {year} {2024})}\BibitemShut {NoStop}%
\bibitem [{\citenamefont {Dong}\ \emph {et~al.}(2018)\citenamefont {Dong},
  \citenamefont {Schumann}, \citenamefont {Hokkanen}, \citenamefont {Chang},
  \citenamefont {Welle}, \citenamefont {Zhou}, \citenamefont {Ras},
  \citenamefont {Xu}, \citenamefont {Wegener},\ and\ \citenamefont
  {Levkin}}]{dong2018superoleophobic}%
  \BibitemOpen
  \bibfield  {author} {\bibinfo {author} {\bibfnamefont {Z.}~\bibnamefont
  {Dong}}, \bibinfo {author} {\bibfnamefont {M.~F.}\ \bibnamefont {Schumann}},
  \bibinfo {author} {\bibfnamefont {M.~J.}\ \bibnamefont {Hokkanen}}, \bibinfo
  {author} {\bibfnamefont {B.}~\bibnamefont {Chang}}, \bibinfo {author}
  {\bibfnamefont {A.}~\bibnamefont {Welle}}, \bibinfo {author} {\bibfnamefont
  {Q.}~\bibnamefont {Zhou}}, \bibinfo {author} {\bibfnamefont {R.~H.}\
  \bibnamefont {Ras}}, \bibinfo {author} {\bibfnamefont {Z.}~\bibnamefont
  {Xu}}, \bibinfo {author} {\bibfnamefont {M.}~\bibnamefont {Wegener}},\ and\
  \bibinfo {author} {\bibfnamefont {P.~A.}\ \bibnamefont {Levkin}},\ }\bibfield
   {title} {\bibinfo {title} {Superoleophobic slippery lubricant-infused
  surfaces: combining two extremes in the same surface},\ }\href@noop {}
  {\bibfield  {journal} {\bibinfo  {journal} {Advanced Materials}\ }\textbf
  {\bibinfo {volume} {30}},\ \bibinfo {pages} {1803890} (\bibinfo {year}
  {2018})}\BibitemShut {NoStop}%
\bibitem [{\citenamefont {{\"O}ner}\ and\ \citenamefont
  {McCarthy}(2000)}]{oner2000ultrahydrophobic}%
  \BibitemOpen
  \bibfield  {author} {\bibinfo {author} {\bibfnamefont {D.}~\bibnamefont
  {{\"O}ner}}\ and\ \bibinfo {author} {\bibfnamefont {T.~J.}\ \bibnamefont
  {McCarthy}},\ }\bibfield  {title} {\bibinfo {title} {Ultrahydrophobic
  surfaces. effects of topography length scales on wettability},\ }\href@noop
  {} {\bibfield  {journal} {\bibinfo  {journal} {Langmuir}\ }\textbf {\bibinfo
  {volume} {16}},\ \bibinfo {pages} {7777} (\bibinfo {year}
  {2000})}\BibitemShut {NoStop}%
\bibitem [{\citenamefont {Priest}\ \emph {et~al.}(2007)\citenamefont {Priest},
  \citenamefont {Sedev},\ and\ \citenamefont {Ralston}}]{priest2007asymmetric}%
  \BibitemOpen
  \bibfield  {author} {\bibinfo {author} {\bibfnamefont {C.}~\bibnamefont
  {Priest}}, \bibinfo {author} {\bibfnamefont {R.}~\bibnamefont {Sedev}},\ and\
  \bibinfo {author} {\bibfnamefont {J.}~\bibnamefont {Ralston}},\ }\bibfield
  {title} {\bibinfo {title} {Asymmetric wetting hysteresis on chemical
  defects},\ }\href@noop {} {\bibfield  {journal} {\bibinfo  {journal}
  {Physical Review Letters}\ }\textbf {\bibinfo {volume} {99}},\ \bibinfo
  {pages} {026103} (\bibinfo {year} {2007})}\BibitemShut {NoStop}%
\bibitem [{\citenamefont {Choi}\ \emph {et~al.}(2009)\citenamefont {Choi},
  \citenamefont {Tuteja}, \citenamefont {Mabry}, \citenamefont {Cohen},\ and\
  \citenamefont {McKinley}}]{choi2009modified}%
  \BibitemOpen
  \bibfield  {author} {\bibinfo {author} {\bibfnamefont {W.}~\bibnamefont
  {Choi}}, \bibinfo {author} {\bibfnamefont {A.}~\bibnamefont {Tuteja}},
  \bibinfo {author} {\bibfnamefont {J.~M.}\ \bibnamefont {Mabry}}, \bibinfo
  {author} {\bibfnamefont {R.~E.}\ \bibnamefont {Cohen}},\ and\ \bibinfo
  {author} {\bibfnamefont {G.~H.}\ \bibnamefont {McKinley}},\ }\bibfield
  {title} {\bibinfo {title} {A modified cassie--baxter relationship to explain
  contact angle hysteresis and anisotropy on non-wetting textured surfaces},\
  }\href@noop {} {\bibfield  {journal} {\bibinfo  {journal} {Journal of Colloid
  and Interface Science}\ }\textbf {\bibinfo {volume} {339}},\ \bibinfo {pages}
  {208} (\bibinfo {year} {2009})}\BibitemShut {NoStop}%
\bibitem [{\citenamefont {Dorrer}\ and\ \citenamefont
  {R{\"u}he}(2007)}]{dorrer2007contact}%
  \BibitemOpen
  \bibfield  {author} {\bibinfo {author} {\bibfnamefont {C.}~\bibnamefont
  {Dorrer}}\ and\ \bibinfo {author} {\bibfnamefont {J.}~\bibnamefont
  {R{\"u}he}},\ }\bibfield  {title} {\bibinfo {title} {Contact line shape on
  ultrahydrophobic post surfaces},\ }\href@noop {} {\bibfield  {journal}
  {\bibinfo  {journal} {Langmuir}\ }\textbf {\bibinfo {volume} {23}},\ \bibinfo
  {pages} {3179} (\bibinfo {year} {2007})}\BibitemShut {NoStop}%
\bibitem [{\citenamefont {Gauthier}\ \emph {et~al.}(2013)\citenamefont
  {Gauthier}, \citenamefont {Rivetti}, \citenamefont {Teisseire},\ and\
  \citenamefont {Barthel}}]{gauthier2013role}%
  \BibitemOpen
  \bibfield  {author} {\bibinfo {author} {\bibfnamefont {A.}~\bibnamefont
  {Gauthier}}, \bibinfo {author} {\bibfnamefont {M.}~\bibnamefont {Rivetti}},
  \bibinfo {author} {\bibfnamefont {J.}~\bibnamefont {Teisseire}},\ and\
  \bibinfo {author} {\bibfnamefont {E.}~\bibnamefont {Barthel}},\ }\bibfield
  {title} {\bibinfo {title} {Role of kinks in the dynamics of contact lines
  receding on superhydrophobic surfaces},\ }\href@noop {} {\bibfield  {journal}
  {\bibinfo  {journal} {Physical Review Letters}\ }\textbf {\bibinfo {volume}
  {110}},\ \bibinfo {pages} {046101} (\bibinfo {year} {2013})}\BibitemShut
  {NoStop}%
\bibitem [{\citenamefont {Dorrer}\ and\ \citenamefont
  {R{\"u}he}(2006)}]{dorrer2006advancing}%
  \BibitemOpen
  \bibfield  {author} {\bibinfo {author} {\bibfnamefont {C.}~\bibnamefont
  {Dorrer}}\ and\ \bibinfo {author} {\bibfnamefont {J.}~\bibnamefont
  {R{\"u}he}},\ }\bibfield  {title} {\bibinfo {title} {Advancing and receding
  motion of droplets on ultrahydrophobic post surfaces},\ }\href@noop {}
  {\bibfield  {journal} {\bibinfo  {journal} {Langmuir}\ }\textbf {\bibinfo
  {volume} {22}},\ \bibinfo {pages} {7652} (\bibinfo {year}
  {2006})}\BibitemShut {NoStop}%
\bibitem [{\citenamefont {Iliev}\ and\ \citenamefont
  {Pesheva}(2016)}]{iliev2016contact}%
  \BibitemOpen
  \bibfield  {author} {\bibinfo {author} {\bibfnamefont {S.}~\bibnamefont
  {Iliev}}\ and\ \bibinfo {author} {\bibfnamefont {N.}~\bibnamefont
  {Pesheva}},\ }\bibfield  {title} {\bibinfo {title} {Contact-angle hysteresis
  on periodic microtextured surfaces: Strongly corrugated liquid interfaces},\
  }\href@noop {} {\bibfield  {journal} {\bibinfo  {journal} {Physical Review
  E}\ }\textbf {\bibinfo {volume} {93}},\ \bibinfo {pages} {062801} (\bibinfo
  {year} {2016})}\BibitemShut {NoStop}%
\bibitem [{\citenamefont {Kusumaatmaja}\ and\ \citenamefont
  {Yeomans}(2007)}]{kusumaatmaja2007modeling}%
  \BibitemOpen
  \bibfield  {author} {\bibinfo {author} {\bibfnamefont {H.}~\bibnamefont
  {Kusumaatmaja}}\ and\ \bibinfo {author} {\bibfnamefont {J.}~\bibnamefont
  {Yeomans}},\ }\bibfield  {title} {\bibinfo {title} {Modeling contact angle
  hysteresis on chemically patterned and superhydrophobic surfaces},\
  }\href@noop {} {\bibfield  {journal} {\bibinfo  {journal} {Langmuir}\
  }\textbf {\bibinfo {volume} {23}},\ \bibinfo {pages} {6019} (\bibinfo {year}
  {2007})}\BibitemShut {NoStop}%
\bibitem [{\citenamefont {Rivetti}\ \emph {et~al.}(2015)\citenamefont
  {Rivetti}, \citenamefont {Teisseire},\ and\ \citenamefont
  {Barthel}}]{rivetti2015surface}%
  \BibitemOpen
  \bibfield  {author} {\bibinfo {author} {\bibfnamefont {M.}~\bibnamefont
  {Rivetti}}, \bibinfo {author} {\bibfnamefont {J.}~\bibnamefont {Teisseire}},\
  and\ \bibinfo {author} {\bibfnamefont {E.}~\bibnamefont {Barthel}},\
  }\bibfield  {title} {\bibinfo {title} {Surface fraction dependence of contact
  angles induced by kinks in the triple line},\ }\href@noop {} {\bibfield
  {journal} {\bibinfo  {journal} {Physical Review Letters}\ }\textbf {\bibinfo
  {volume} {115}},\ \bibinfo {pages} {016101} (\bibinfo {year}
  {2015})}\BibitemShut {NoStop}%
\bibitem [{\citenamefont {Mognetti}\ and\ \citenamefont
  {Yeomans}(2010)}]{mognetti2010modeling}%
  \BibitemOpen
  \bibfield  {author} {\bibinfo {author} {\bibfnamefont {B.~M.}\ \bibnamefont
  {Mognetti}}\ and\ \bibinfo {author} {\bibfnamefont {J.}~\bibnamefont
  {Yeomans}},\ }\bibfield  {title} {\bibinfo {title} {Modeling receding contact
  lines on superhydrophobic surfaces},\ }\href@noop {} {\bibfield  {journal}
  {\bibinfo  {journal} {Langmuir}\ }\textbf {\bibinfo {volume} {26}},\ \bibinfo
  {pages} {18162} (\bibinfo {year} {2010})}\BibitemShut {NoStop}%
\bibitem [{\citenamefont {Priest}\ \emph {et~al.}(2009)\citenamefont {Priest},
  \citenamefont {Albrecht}, \citenamefont {Sedev},\ and\ \citenamefont
  {Ralston}}]{priest2009asymmetric}%
  \BibitemOpen
  \bibfield  {author} {\bibinfo {author} {\bibfnamefont {C.}~\bibnamefont
  {Priest}}, \bibinfo {author} {\bibfnamefont {T.~W.}\ \bibnamefont
  {Albrecht}}, \bibinfo {author} {\bibfnamefont {R.}~\bibnamefont {Sedev}},\
  and\ \bibinfo {author} {\bibfnamefont {J.}~\bibnamefont {Ralston}},\
  }\bibfield  {title} {\bibinfo {title} {Asymmetric wetting hysteresis on
  hydrophobic microstructured surfaces},\ }\href@noop {} {\bibfield  {journal}
  {\bibinfo  {journal} {Langmuir}\ }\textbf {\bibinfo {volume} {25}},\ \bibinfo
  {pages} {5655} (\bibinfo {year} {2009})}\BibitemShut {NoStop}%
\bibitem [{\citenamefont {Reyssat}\ and\ \citenamefont
  {Qu{\'e}r{\'e}}(2009)}]{reyssat2009contact}%
  \BibitemOpen
  \bibfield  {author} {\bibinfo {author} {\bibfnamefont {M.}~\bibnamefont
  {Reyssat}}\ and\ \bibinfo {author} {\bibfnamefont {D.}~\bibnamefont
  {Qu{\'e}r{\'e}}},\ }\bibfield  {title} {\bibinfo {title} {Contact angle
  hysteresis generated by strong dilute defects},\ }\href@noop {} {\bibfield
  {journal} {\bibinfo  {journal} {The Journal of Physical Chemistry B}\
  }\textbf {\bibinfo {volume} {113}},\ \bibinfo {pages} {3906} (\bibinfo {year}
  {2009})}\BibitemShut {NoStop}%
\bibitem [{\citenamefont {Yeh}\ \emph {et~al.}(2008)\citenamefont {Yeh},
  \citenamefont {Chen},\ and\ \citenamefont {Chang}}]{yeh2008contact}%
  \BibitemOpen
  \bibfield  {author} {\bibinfo {author} {\bibfnamefont {K.-Y.}\ \bibnamefont
  {Yeh}}, \bibinfo {author} {\bibfnamefont {L.-J.}\ \bibnamefont {Chen}},\ and\
  \bibinfo {author} {\bibfnamefont {J.-Y.}\ \bibnamefont {Chang}},\ }\bibfield
  {title} {\bibinfo {title} {Contact angle hysteresis on regular pillar-like
  hydrophobic surfaces},\ }\href@noop {} {\bibfield  {journal} {\bibinfo
  {journal} {Langmuir}\ }\textbf {\bibinfo {volume} {24}},\ \bibinfo {pages}
  {245} (\bibinfo {year} {2008})}\BibitemShut {NoStop}%
\bibitem [{\citenamefont {Kwon}\ \emph {et~al.}(2010)\citenamefont {Kwon},
  \citenamefont {Choi}, \citenamefont {Anantharaju}, \citenamefont {Lee},
  \citenamefont {Panchagnula},\ and\ \citenamefont
  {Patankar}}]{kwon2010cassie}%
  \BibitemOpen
  \bibfield  {author} {\bibinfo {author} {\bibfnamefont {Y.}~\bibnamefont
  {Kwon}}, \bibinfo {author} {\bibfnamefont {S.}~\bibnamefont {Choi}}, \bibinfo
  {author} {\bibfnamefont {N.}~\bibnamefont {Anantharaju}}, \bibinfo {author}
  {\bibfnamefont {J.}~\bibnamefont {Lee}}, \bibinfo {author} {\bibfnamefont
  {M.}~\bibnamefont {Panchagnula}},\ and\ \bibinfo {author} {\bibfnamefont
  {N.}~\bibnamefont {Patankar}},\ }\bibfield  {title} {\bibinfo {title} {Is the
  cassie- baxter formula relevant?},\ }\href@noop {} {\bibfield  {journal}
  {\bibinfo  {journal} {Langmuir}\ }\textbf {\bibinfo {volume} {26}},\ \bibinfo
  {pages} {17528} (\bibinfo {year} {2010})}\BibitemShut {NoStop}%
\bibitem [{\citenamefont {Schellenberger}\ \emph {et~al.}(2016)\citenamefont
  {Schellenberger}, \citenamefont {Encinas}, \citenamefont {Vollmer},\ and\
  \citenamefont {Butt}}]{schellenberger2016water}%
  \BibitemOpen
  \bibfield  {author} {\bibinfo {author} {\bibfnamefont {F.}~\bibnamefont
  {Schellenberger}}, \bibinfo {author} {\bibfnamefont {N.}~\bibnamefont
  {Encinas}}, \bibinfo {author} {\bibfnamefont {D.}~\bibnamefont {Vollmer}},\
  and\ \bibinfo {author} {\bibfnamefont {H.-J.}\ \bibnamefont {Butt}},\
  }\bibfield  {title} {\bibinfo {title} {How water advances on superhydrophobic
  surfaces},\ }\href@noop {} {\bibfield  {journal} {\bibinfo  {journal}
  {Physical Review Letters}\ }\textbf {\bibinfo {volume} {116}},\ \bibinfo
  {pages} {096101} (\bibinfo {year} {2016})}\BibitemShut {NoStop}%
\bibitem [{\citenamefont {Butt}(2008)}]{butt2008capillary}%
  \BibitemOpen
  \bibfield  {author} {\bibinfo {author} {\bibfnamefont {H.-J.}\ \bibnamefont
  {Butt}},\ }\bibfield  {title} {\bibinfo {title} {Capillary forces: Influence
  of roughness and heterogeneity},\ }\href@noop {} {\bibfield  {journal}
  {\bibinfo  {journal} {Langmuir}\ }\textbf {\bibinfo {volume} {24}},\ \bibinfo
  {pages} {4715} (\bibinfo {year} {2008})}\BibitemShut {NoStop}%
\bibitem [{\citenamefont {Baret}\ and\ \citenamefont
  {Brinkmann}(2006)}]{baret2006wettability}%
  \BibitemOpen
  \bibfield  {author} {\bibinfo {author} {\bibfnamefont {J.-C.}\ \bibnamefont
  {Baret}}\ and\ \bibinfo {author} {\bibfnamefont {M.}~\bibnamefont
  {Brinkmann}},\ }\bibfield  {title} {\bibinfo {title} {Wettability control of
  droplet deposition and detachment},\ }\href@noop {} {\bibfield  {journal}
  {\bibinfo  {journal} {Physical Review Letters}\ }\textbf {\bibinfo {volume}
  {96}},\ \bibinfo {pages} {146106} (\bibinfo {year} {2006})}\BibitemShut
  {NoStop}%
\bibitem [{\citenamefont {Daniel}\ and\ \citenamefont
  {Koh}(2023)}]{daniel2023droplet}%
  \BibitemOpen
  \bibfield  {author} {\bibinfo {author} {\bibfnamefont {D.}~\bibnamefont
  {Daniel}}\ and\ \bibinfo {author} {\bibfnamefont {X.~Q.}\ \bibnamefont
  {Koh}},\ }\bibfield  {title} {\bibinfo {title} {Droplet detachment force and
  its relation to young--dupre adhesion},\ }\href@noop {} {\bibfield  {journal}
  {\bibinfo  {journal} {Soft Matter}\ }\textbf {\bibinfo {volume} {19}},\
  \bibinfo {pages} {8434} (\bibinfo {year} {2023})}\BibitemShut {NoStop}%
\bibitem [{\citenamefont {Sadullah}\ \emph {et~al.}(2024)\citenamefont
  {Sadullah}, \citenamefont {Xu}, \citenamefont {Arunachalam},\ and\
  \citenamefont {Mishra}}]{sadullah2024predicting}%
  \BibitemOpen
  \bibfield  {author} {\bibinfo {author} {\bibfnamefont {M.~S.}\ \bibnamefont
  {Sadullah}}, \bibinfo {author} {\bibfnamefont {Y.}~\bibnamefont {Xu}},
  \bibinfo {author} {\bibfnamefont {S.}~\bibnamefont {Arunachalam}},\ and\
  \bibinfo {author} {\bibfnamefont {H.}~\bibnamefont {Mishra}},\ }\bibfield
  {title} {\bibinfo {title} {Predicting droplet detachment force:
  Young-dupr{\'e} model fails, young-laplace model prevails},\ }\href@noop {}
  {\bibfield  {journal} {\bibinfo  {journal} {Communications Physics}\ }\textbf
  {\bibinfo {volume} {7}},\ \bibinfo {pages} {89} (\bibinfo {year}
  {2024})}\BibitemShut {NoStop}%
\bibitem [{\citenamefont {Tadmor}\ \emph {et~al.}(2008)\citenamefont {Tadmor},
  \citenamefont {Dang}, \citenamefont {Leh}, \citenamefont {Bahadur},\ and\
  \citenamefont {Chaurasia}}]{tadmor2008centrifugal}%
  \BibitemOpen
  \bibfield  {author} {\bibinfo {author} {\bibfnamefont {R.}~\bibnamefont
  {Tadmor}}, \bibinfo {author} {\bibfnamefont {L.}~\bibnamefont {Dang}},
  \bibinfo {author} {\bibfnamefont {A.}~\bibnamefont {Leh}}, \bibinfo {author}
  {\bibfnamefont {P.}~\bibnamefont {Bahadur}},\ and\ \bibinfo {author}
  {\bibfnamefont {K.}~\bibnamefont {Chaurasia}},\ }\bibfield  {title} {\bibinfo
  {title} {Centrifugal adhesion balance (\uppercase{CAB}): A novel surface
  characterization technique},\ }in\ \href@noop {} {\emph {\bibinfo {booktitle}
  {APS March Meeting Abstracts}}}\ (\bibinfo {year} {2008})\ pp.\ \bibinfo
  {pages} {Q22--013}\BibitemShut {NoStop}%
\bibitem [{\citenamefont {Chen}\ \emph {et~al.}(2017)\citenamefont {Chen},
  \citenamefont {Zhang}, \citenamefont {Gao}, \citenamefont {Liu},\ and\
  \citenamefont {Zhang}}]{chen2017direction}%
  \BibitemOpen
  \bibfield  {author} {\bibinfo {author} {\bibfnamefont {S.}~\bibnamefont
  {Chen}}, \bibinfo {author} {\bibfnamefont {B.}~\bibnamefont {Zhang}},
  \bibinfo {author} {\bibfnamefont {X.}~\bibnamefont {Gao}}, \bibinfo {author}
  {\bibfnamefont {Z.}~\bibnamefont {Liu}},\ and\ \bibinfo {author}
  {\bibfnamefont {X.}~\bibnamefont {Zhang}},\ }\bibfield  {title} {\bibinfo
  {title} {Direction dependence of adhesion force for droplets on rough
  substrates},\ }\href@noop {} {\bibfield  {journal} {\bibinfo  {journal}
  {Langmuir}\ }\textbf {\bibinfo {volume} {33}},\ \bibinfo {pages} {2472}
  (\bibinfo {year} {2017})}\BibitemShut {NoStop}%
\bibitem [{\citenamefont {Sudersan}\ \emph {et~al.}(2023)\citenamefont
  {Sudersan}, \citenamefont {M{\"u}ller}, \citenamefont {Hormozi},
  \citenamefont {Li}, \citenamefont {Butt},\ and\ \citenamefont
  {Kappl}}]{sudersan2023method}%
  \BibitemOpen
  \bibfield  {author} {\bibinfo {author} {\bibfnamefont {P.}~\bibnamefont
  {Sudersan}}, \bibinfo {author} {\bibfnamefont {M.}~\bibnamefont
  {M{\"u}ller}}, \bibinfo {author} {\bibfnamefont {M.}~\bibnamefont {Hormozi}},
  \bibinfo {author} {\bibfnamefont {S.}~\bibnamefont {Li}}, \bibinfo {author}
  {\bibfnamefont {H.-J.}\ \bibnamefont {Butt}},\ and\ \bibinfo {author}
  {\bibfnamefont {M.}~\bibnamefont {Kappl}},\ }\bibfield  {title} {\bibinfo
  {title} {Method to measure surface tension of microdroplets using standard
  afm cantilever tips},\ }\href@noop {} {\bibfield  {journal} {\bibinfo
  {journal} {Langmuir}\ }\textbf {\bibinfo {volume} {39}},\ \bibinfo {pages}
  {10367} (\bibinfo {year} {2023})}\BibitemShut {NoStop}%
\bibitem [{\citenamefont {Kumar}\ and\ \citenamefont
  {Harvie}(2024)}]{kumar2024numerical}%
  \BibitemOpen
  \bibfield  {author} {\bibinfo {author} {\bibfnamefont {P.}~\bibnamefont
  {Kumar}}\ and\ \bibinfo {author} {\bibfnamefont {D.~J.~E.}\ \bibnamefont
  {Harvie}},\ }\bibfield  {title} {\bibinfo {title} {Energy dissipation during
  wenzel wetting via roughness scale interface dynamics},\ }\href@noop {}
  {\bibfield  {journal} {\bibinfo  {journal} {Langmuir}\ }\textbf {\bibinfo
  {volume} {40}},\ \bibinfo {pages} {16190} (\bibinfo {year}
  {2024})}\BibitemShut {NoStop}%
\bibitem [{\citenamefont {Kumar}\ \emph {et~al.}(2024)\citenamefont {Kumar},
  \citenamefont {Mulvaney},\ and\ \citenamefont
  {Harvie}}]{kumar2024experimental}%
  \BibitemOpen
  \bibfield  {author} {\bibinfo {author} {\bibfnamefont {P.}~\bibnamefont
  {Kumar}}, \bibinfo {author} {\bibfnamefont {P.}~\bibnamefont {Mulvaney}},\
  and\ \bibinfo {author} {\bibfnamefont {D.~J.}\ \bibnamefont {Harvie}},\
  }\bibfield  {title} {\bibinfo {title} {Energy dissipation during homogeneous
  wetting of surfaces with randomly and periodically distributed cylindrical
  pillars},\ }\href@noop {} {\bibfield  {journal} {\bibinfo  {journal} {Journal
  of Colloid and Interface Science}\ }\textbf {\bibinfo {volume} {659}},\
  \bibinfo {pages} {105} (\bibinfo {year} {2024})}\BibitemShut {NoStop}%
\bibitem [{\citenamefont {Raj}\ \emph {et~al.}(2014)\citenamefont {Raj},
  \citenamefont {Adera}, \citenamefont {Enright},\ and\ \citenamefont
  {Wang}}]{raj2014high}%
  \BibitemOpen
  \bibfield  {author} {\bibinfo {author} {\bibfnamefont {R.}~\bibnamefont
  {Raj}}, \bibinfo {author} {\bibfnamefont {S.}~\bibnamefont {Adera}}, \bibinfo
  {author} {\bibfnamefont {R.}~\bibnamefont {Enright}},\ and\ \bibinfo {author}
  {\bibfnamefont {E.~N.}\ \bibnamefont {Wang}},\ }\bibfield  {title} {\bibinfo
  {title} {High-resolution liquid patterns via three-dimensional droplet shape
  control},\ }\href@noop {} {\bibfield  {journal} {\bibinfo  {journal} {Nature
  communications}\ }\textbf {\bibinfo {volume} {5}},\ \bibinfo {pages} {4975}
  (\bibinfo {year} {2014})}\BibitemShut {NoStop}%
\bibitem [{\citenamefont {He}\ \emph {et~al.}(2004)\citenamefont {He},
  \citenamefont {Lee},\ and\ \citenamefont {Patankar}}]{he2004contact}%
  \BibitemOpen
  \bibfield  {author} {\bibinfo {author} {\bibfnamefont {B.}~\bibnamefont
  {He}}, \bibinfo {author} {\bibfnamefont {J.}~\bibnamefont {Lee}},\ and\
  \bibinfo {author} {\bibfnamefont {N.~A.}\ \bibnamefont {Patankar}},\
  }\bibfield  {title} {\bibinfo {title} {Contact angle hysteresis on rough
  hydrophobic surfaces},\ }\href@noop {} {\bibfield  {journal} {\bibinfo
  {journal} {Colloids and Surfaces A: Physicochemical and Engineering Aspects}\
  }\textbf {\bibinfo {volume} {248}},\ \bibinfo {pages} {101} (\bibinfo {year}
  {2004})}\BibitemShut {NoStop}%
\bibitem [{\citenamefont {Darvish}\ \emph {et~al.}(2026)\citenamefont
  {Darvish}, \citenamefont {Isaacs}, \citenamefont {Shumaly}, \citenamefont
  {Delance},\ and\ \citenamefont {Butt}}]{darvish2026water}%
  \BibitemOpen
  \bibfield  {author} {\bibinfo {author} {\bibfnamefont {F.}~\bibnamefont
  {Darvish}}, \bibinfo {author} {\bibfnamefont {M.}~\bibnamefont {Isaacs}},
  \bibinfo {author} {\bibfnamefont {S.}~\bibnamefont {Shumaly}}, \bibinfo
  {author} {\bibfnamefont {L.}~\bibnamefont {Delance}},\ and\ \bibinfo {author}
  {\bibfnamefont {H.-J.}\ \bibnamefont {Butt}},\ }\bibfield  {title} {\bibinfo
  {title} {Water drops sliding over arrays of janus micropillars with
  hydrophilic tops: A new mechanism of drop charging},\ }\href@noop {}
  {\bibfield  {journal} {\bibinfo  {journal} {Small}\ ,\ \bibinfo {pages}
  {e11728}} (\bibinfo {year} {2026})}\BibitemShut {NoStop}%
\bibitem [{\citenamefont {de~Gennes}\ \emph {et~al.}(2003)\citenamefont
  {de~Gennes}, \citenamefont {Brochard~Wyart},\ and\ \citenamefont
  {Qu{\'e}r{\'e}}}]{de2003capillarity}%
  \BibitemOpen
  \bibfield  {author} {\bibinfo {author} {\bibfnamefont {P.~G.}\ \bibnamefont
  {de~Gennes}}, \bibinfo {author} {\bibfnamefont {F.}~\bibnamefont
  {Brochard~Wyart}},\ and\ \bibinfo {author} {\bibfnamefont {D.}~\bibnamefont
  {Qu{\'e}r{\'e}}},\ }\href@noop {} {\emph {\bibinfo {title} {Capillarity and
  wetting phenomena: drops, bubbles, pearls, waves}}}\ (\bibinfo  {publisher}
  {Springer Science \& Business Media},\ \bibinfo {year} {2003})\BibitemShut
  {NoStop}%
\bibitem [{\citenamefont {Gibbs}(1961)}]{Gibbs1961}%
  \BibitemOpen
  \bibfield  {author} {\bibinfo {author} {\bibfnamefont {J.~W.}\ \bibnamefont
  {Gibbs}},\ }\href {https://books.google.com/books?id=WlYLygEACAAJ} {\emph
  {\bibinfo {title} {The Scientific Papers of J. Willard Gibbs, Vol. I:
  Thermodynamics}}}\ (\bibinfo  {publisher} {Dover Publications},\ \bibinfo
  {year} {1961})\BibitemShut {NoStop}%
\bibitem [{\citenamefont {Paxson}\ and\ \citenamefont
  {Varanasi}(2013)}]{paxson2013self}%
  \BibitemOpen
  \bibfield  {author} {\bibinfo {author} {\bibfnamefont {A.~T.}\ \bibnamefont
  {Paxson}}\ and\ \bibinfo {author} {\bibfnamefont {K.~K.}\ \bibnamefont
  {Varanasi}},\ }\bibfield  {title} {\bibinfo {title} {Self-similarity of
  contact line depinning from textured surfaces},\ }\href@noop {} {\bibfield
  {journal} {\bibinfo  {journal} {Nature Communications}\ }\textbf {\bibinfo
  {volume} {4}},\ \bibinfo {pages} {1492} (\bibinfo {year} {2013})}\BibitemShut
  {NoStop}%
\bibitem [{\citenamefont {Dyson}(1988)}]{dyson1988contact}%
  \BibitemOpen
  \bibfield  {author} {\bibinfo {author} {\bibfnamefont {D.}~\bibnamefont
  {Dyson}},\ }\bibfield  {title} {\bibinfo {title} {Contact line stability at
  edges: Comments on gibbs’s inequalities},\ }\href@noop {} {\bibfield
  {journal} {\bibinfo  {journal} {The Physics of Fluids}\ }\textbf {\bibinfo
  {volume} {31}},\ \bibinfo {pages} {229} (\bibinfo {year} {1988})}\BibitemShut
  {NoStop}%
\bibitem [{\citenamefont {Gao}\ and\ \citenamefont
  {McCarthy}(2006)}]{gao2006lotus}%
  \BibitemOpen
  \bibfield  {author} {\bibinfo {author} {\bibfnamefont {L.}~\bibnamefont
  {Gao}}\ and\ \bibinfo {author} {\bibfnamefont {T.~J.}\ \bibnamefont
  {McCarthy}},\ }\bibfield  {title} {\bibinfo {title} {The “lotus effect”
  explained: two reasons why two length scales of topography are important},\
  }\href@noop {} {\bibfield  {journal} {\bibinfo  {journal} {Langmuir}\
  }\textbf {\bibinfo {volume} {22}},\ \bibinfo {pages} {2966} (\bibinfo {year}
  {2006})}\BibitemShut {NoStop}%
\bibitem [{\citenamefont {Dorrer}\ and\ \citenamefont
  {R{\"u}he}(2011)}]{dorrer2011micro}%
  \BibitemOpen
  \bibfield  {author} {\bibinfo {author} {\bibfnamefont {C.}~\bibnamefont
  {Dorrer}}\ and\ \bibinfo {author} {\bibfnamefont {J.}~\bibnamefont
  {R{\"u}he}},\ }\bibfield  {title} {\bibinfo {title} {Micro to nano: Surface
  size scale and superhydrophobicity},\ }\href@noop {} {\bibfield  {journal}
  {\bibinfo  {journal} {Beilstein Journal of Nanotechnology}\ }\textbf
  {\bibinfo {volume} {2}},\ \bibinfo {pages} {327} (\bibinfo {year}
  {2011})}\BibitemShut {NoStop}%
\bibitem [{\citenamefont {Yeong}\ \emph {et~al.}(2015)\citenamefont {Yeong},
  \citenamefont {Milionis}, \citenamefont {Loth},\ and\ \citenamefont
  {Bayer}}]{yeong2015microscopic}%
  \BibitemOpen
  \bibfield  {author} {\bibinfo {author} {\bibfnamefont {Y.~H.}\ \bibnamefont
  {Yeong}}, \bibinfo {author} {\bibfnamefont {A.}~\bibnamefont {Milionis}},
  \bibinfo {author} {\bibfnamefont {E.}~\bibnamefont {Loth}},\ and\ \bibinfo
  {author} {\bibfnamefont {I.~S.}\ \bibnamefont {Bayer}},\ }\bibfield  {title}
  {\bibinfo {title} {Microscopic receding contact line dynamics on pillar and
  irregular superhydrophobic surfaces},\ }\href@noop {} {\bibfield  {journal}
  {\bibinfo  {journal} {Scientific Reports}\ }\textbf {\bibinfo {volume} {5}},\
  \bibinfo {pages} {8384} (\bibinfo {year} {2015})}\BibitemShut {NoStop}%
\bibitem [{\citenamefont {Nguyen}\ \emph {et~al.}(2016)\citenamefont {Nguyen},
  \citenamefont {Tsukagoshi}, \citenamefont {Takahashi}, \citenamefont
  {Matsumoto},\ and\ \citenamefont {Shimoyama}}]{nguyen2016depinning}%
  \BibitemOpen
  \bibfield  {author} {\bibinfo {author} {\bibfnamefont {T.-V.}\ \bibnamefont
  {Nguyen}}, \bibinfo {author} {\bibfnamefont {T.}~\bibnamefont {Tsukagoshi}},
  \bibinfo {author} {\bibfnamefont {H.}~\bibnamefont {Takahashi}}, \bibinfo
  {author} {\bibfnamefont {K.}~\bibnamefont {Matsumoto}},\ and\ \bibinfo
  {author} {\bibfnamefont {I.}~\bibnamefont {Shimoyama}},\ }\bibfield  {title}
  {\bibinfo {title} {Depinning-induced capillary wave during the sliding of a
  droplet on a textured surface},\ }\href@noop {} {\bibfield  {journal}
  {\bibinfo  {journal} {Langmuir}\ }\textbf {\bibinfo {volume} {32}},\ \bibinfo
  {pages} {9523} (\bibinfo {year} {2016})}\BibitemShut {NoStop}%
\bibitem [{\citenamefont {Brakke}(1992)}]{brakke1992software}%
  \BibitemOpen
  \bibfield  {author} {\bibinfo {author} {\bibfnamefont {K.~A.}\ \bibnamefont
  {Brakke}},\ }\bibfield  {title} {\bibinfo {title} {The surface evolver},\
  }\href {https://doi.org/10.1080/10586458.1992.10504253} {\bibfield  {journal}
  {\bibinfo  {journal} {Experimental Mathematics}\ }\textbf {\bibinfo {volume}
  {1}},\ \bibinfo {pages} {141} (\bibinfo {year} {1992})}\BibitemShut {NoStop}%
\bibitem [{\citenamefont {Brakke}(2013)}]{brakke1994manual}%
  \BibitemOpen
  \bibfield  {author} {\bibinfo {author} {\bibfnamefont {K.~A.}\ \bibnamefont
  {Brakke}},\ }\href {https://kenbrakke.com/evolver/downloads/manual270.pdf}
  {\emph {\bibinfo {title} {Surface Evolver Manual}}},\ \bibinfo {edition}
  {version 2.70}\ ed.\ (\bibinfo  {publisher} {Mathematics Department,
  Susquehanna University},\ \bibinfo {address} {Selinsgrove, PA},\ \bibinfo
  {year} {2013})\BibitemShut {NoStop}%
\bibitem [{\citenamefont {Barr}(1981)}]{barr1981superquadrics}%
  \BibitemOpen
  \bibfield  {author} {\bibinfo {author} {\bibfnamefont {A.~H.}\ \bibnamefont
  {Barr}},\ }\bibfield  {title} {\bibinfo {title} {Superquadrics and
  angle-preserving transformations},\ }\href@noop {} {\bibfield  {journal}
  {\bibinfo  {journal} {IEEE Computer Graphics and Applications}\ }\textbf
  {\bibinfo {volume} {1}},\ \bibinfo {pages} {11} (\bibinfo {year}
  {1981})}\BibitemShut {NoStop}%
\bibitem [{Sup()}]{SupplementalMaterial}%
  \BibitemOpen
  \href@noop {} {}\bibinfo {note} {See Supplemental Material at [URL] for
  detailed information on the numerical method used to calculate adhesion
  forces, a flowchart of the proposed numerical method, simulation animations,
  variations in droplet pressure during the receding stage in evaporation mode,
  experimental videos, and variations in the scale factor during a typical
  simulation.}\BibitemShut {Stop}%
\bibitem [{\citenamefont {Wong}\ \emph {et~al.}(2020)\citenamefont {Wong},
  \citenamefont {Corrales}, \citenamefont {Naga}, \citenamefont {Baumli},
  \citenamefont {Kaltbeitzel}, \citenamefont {Kappl}, \citenamefont
  {Papadopoulos}, \citenamefont {Vollmer},\ and\ \citenamefont
  {Butt}}]{wong2020microdroplet}%
  \BibitemOpen
  \bibfield  {author} {\bibinfo {author} {\bibfnamefont {W.~S.}\ \bibnamefont
  {Wong}}, \bibinfo {author} {\bibfnamefont {T.~P.}\ \bibnamefont {Corrales}},
  \bibinfo {author} {\bibfnamefont {A.}~\bibnamefont {Naga}}, \bibinfo {author}
  {\bibfnamefont {P.}~\bibnamefont {Baumli}}, \bibinfo {author} {\bibfnamefont
  {A.}~\bibnamefont {Kaltbeitzel}}, \bibinfo {author} {\bibfnamefont
  {M.}~\bibnamefont {Kappl}}, \bibinfo {author} {\bibfnamefont
  {P.}~\bibnamefont {Papadopoulos}}, \bibinfo {author} {\bibfnamefont
  {D.}~\bibnamefont {Vollmer}},\ and\ \bibinfo {author} {\bibfnamefont {H.-J.}\
  \bibnamefont {Butt}},\ }\bibfield  {title} {\bibinfo {title} {Microdroplet
  contaminants: when and why superamphiphobic surfaces are not self-cleaning},\
  }\href@noop {} {\bibfield  {journal} {\bibinfo  {journal} {ACS Nano}\
  }\textbf {\bibinfo {volume} {14}},\ \bibinfo {pages} {3836} (\bibinfo {year}
  {2020})}\BibitemShut {NoStop}%
\bibitem [{\citenamefont {Papadopoulos}\ \emph {et~al.}(2013)\citenamefont
  {Papadopoulos}, \citenamefont {Mammen}, \citenamefont {Deng}, \citenamefont
  {Vollmer},\ and\ \citenamefont {Butt}}]{papadopoulos2013superhydrophobicity}%
  \BibitemOpen
  \bibfield  {author} {\bibinfo {author} {\bibfnamefont {P.}~\bibnamefont
  {Papadopoulos}}, \bibinfo {author} {\bibfnamefont {L.}~\bibnamefont
  {Mammen}}, \bibinfo {author} {\bibfnamefont {X.}~\bibnamefont {Deng}},
  \bibinfo {author} {\bibfnamefont {D.}~\bibnamefont {Vollmer}},\ and\ \bibinfo
  {author} {\bibfnamefont {H.-J.}\ \bibnamefont {Butt}},\ }\bibfield  {title}
  {\bibinfo {title} {How superhydrophobicity breaks down},\ }\href@noop {}
  {\bibfield  {journal} {\bibinfo  {journal} {Proceedings of the National
  Academy of Sciences}\ }\textbf {\bibinfo {volume} {110}},\ \bibinfo {pages}
  {3254} (\bibinfo {year} {2013})}\BibitemShut {NoStop}%
\bibitem [{Kum()}]{Kumar2026Code}%
  \BibitemOpen
  \href@noop {} {}\bibinfo {note} {The Surface Evolver code developed for
  simulating droplet adhesion measurements is publicly available at
  \url{https://github.com/jdber1/drop-adhesion-SE}.}\BibitemShut {Stop}%
\bibitem [{\citenamefont {Harvie}(2024)}]{Harvie_2024}%
  \BibitemOpen
  \bibfield  {author} {\bibinfo {author} {\bibfnamefont {D.~J.}\ \bibnamefont
  {Harvie}},\ }\bibfield  {title} {\bibinfo {title} {Contact-angle hysteresis
  on rough surfaces: mechanical energy balance framework},\ }\href
  {https://doi.org/10.1017/jfm.2024.317} {\bibfield  {journal} {\bibinfo
  {journal} {Journal of Fluid Mechanics}\ }\textbf {\bibinfo {volume} {986}},\
  \bibinfo {pages} {A17} (\bibinfo {year} {2024})}\BibitemShut {NoStop}%
\bibitem [{\citenamefont {Jiang}\ \emph {et~al.}(2019)\citenamefont {Jiang},
  \citenamefont {Xu}, \citenamefont {Sarshar},\ and\ \citenamefont
  {Choi}}]{jiang2019generalized}%
  \BibitemOpen
  \bibfield  {author} {\bibinfo {author} {\bibfnamefont {Y.}~\bibnamefont
  {Jiang}}, \bibinfo {author} {\bibfnamefont {W.}~\bibnamefont {Xu}}, \bibinfo
  {author} {\bibfnamefont {M.~A.}\ \bibnamefont {Sarshar}},\ and\ \bibinfo
  {author} {\bibfnamefont {C.-H.}\ \bibnamefont {Choi}},\ }\bibfield  {title}
  {\bibinfo {title} {Generalized models for advancing and receding contact
  angles of fakir droplets on pillared and pored surfaces},\ }\href@noop {}
  {\bibfield  {journal} {\bibinfo  {journal} {Journal of Colloid and Interface
  Science}\ }\textbf {\bibinfo {volume} {552}},\ \bibinfo {pages} {359}
  (\bibinfo {year} {2019})}\BibitemShut {NoStop}%
\bibitem [{\citenamefont {Patankar}(2010)}]{patankar2010hysteresis}%
  \BibitemOpen
  \bibfield  {author} {\bibinfo {author} {\bibfnamefont {N.~A.}\ \bibnamefont
  {Patankar}},\ }\bibfield  {title} {\bibinfo {title} {Hysteresis with regard
  to cassie and wenzel states on superhydrophobic surfaces},\ }\href@noop {}
  {\bibfield  {journal} {\bibinfo  {journal} {Langmuir}\ }\textbf {\bibinfo
  {volume} {26}},\ \bibinfo {pages} {7498} (\bibinfo {year}
  {2010})}\BibitemShut {NoStop}%
\bibitem [{\citenamefont {Liu}\ \emph {et~al.}(2015)\citenamefont {Liu},
  \citenamefont {Chen},\ and\ \citenamefont {Kim}}]{liu2015dynamic}%
  \BibitemOpen
  \bibfield  {author} {\bibinfo {author} {\bibfnamefont {T.~L.}\ \bibnamefont
  {Liu}}, \bibinfo {author} {\bibfnamefont {Z.}~\bibnamefont {Chen}},\ and\
  \bibinfo {author} {\bibfnamefont {C.-J.}\ \bibnamefont {Kim}},\ }\bibfield
  {title} {\bibinfo {title} {A dynamic cassie--baxter model},\ }\href@noop {}
  {\bibfield  {journal} {\bibinfo  {journal} {Soft Matter}\ }\textbf {\bibinfo
  {volume} {11}},\ \bibinfo {pages} {1589} (\bibinfo {year}
  {2015})}\BibitemShut {NoStop}%
\bibitem [{\citenamefont {Smith}\ \emph {et~al.}(2018)\citenamefont {Smith},
  \citenamefont {Mahelona},\ and\ \citenamefont {Hendy}}]{smith2018rolling}%
  \BibitemOpen
  \bibfield  {author} {\bibinfo {author} {\bibfnamefont {A.~F.}\ \bibnamefont
  {Smith}}, \bibinfo {author} {\bibfnamefont {K.}~\bibnamefont {Mahelona}},\
  and\ \bibinfo {author} {\bibfnamefont {S.~C.}\ \bibnamefont {Hendy}},\
  }\bibfield  {title} {\bibinfo {title} {Rolling and slipping of droplets on
  superhydrophobic surfaces},\ }\href@noop {} {\bibfield  {journal} {\bibinfo
  {journal} {Physical Review E}\ }\textbf {\bibinfo {volume} {98}},\ \bibinfo
  {pages} {033113} (\bibinfo {year} {2018})}\BibitemShut {NoStop}%
\bibitem [{\citenamefont {Dekker}\ \emph {et~al.}(2026)\citenamefont {Dekker},
  \citenamefont {Diddens}, \citenamefont {van~der Linden},\ and\ \citenamefont
  {Lohse}}]{dekker2025hidden}%
  \BibitemOpen
  \bibfield  {author} {\bibinfo {author} {\bibfnamefont {P.~J.}\ \bibnamefont
  {Dekker}}, \bibinfo {author} {\bibfnamefont {C.}~\bibnamefont {Diddens}},
  \bibinfo {author} {\bibfnamefont {M.~N.}\ \bibnamefont {van~der Linden}},\
  and\ \bibinfo {author} {\bibfnamefont {D.}~\bibnamefont {Lohse}},\ }\bibfield
   {title} {\bibinfo {title} {Hidden in plain sight: How evaporation impacts
  the pendant drop method},\ }\href@noop {} {\bibfield  {journal} {\bibinfo
  {journal} {Physical Review Fluids}\ }\textbf {\bibinfo {volume} {11}},\
  \bibinfo {pages} {094901} (\bibinfo {year} {2026})}\BibitemShut {NoStop}%
\bibitem [{\citenamefont {Kumar}\ and\ \citenamefont
  {Berry}(2026)}]{Kumar2026Figshare}%
  \BibitemOpen
  \bibfield  {author} {\bibinfo {author} {\bibfnamefont {P.}~\bibnamefont
  {Kumar}}\ and\ \bibinfo {author} {\bibfnamefont {J.~D.}\ \bibnamefont
  {Berry}},\ }\href@noop {} {\bibinfo {title} {Numerical and experimental
  data}} (\bibinfo {year} {2026}),\ \bibinfo {note} {\MakeUppercase{F}igshare,
  \url{https://doi.org/10.26188/34746276}}\BibitemShut {NoStop}%
\end{thebibliography}%

\end{document}